\documentclass[aps,pre,floatfix,reprint,groupedaddress,showkeys,superscriptaddress]{revtex4-1}
\usepackage[utf8x]{inputenc}
\usepackage{graphicx}
\usepackage{dcolumn}
\usepackage{bm}
\usepackage{epstopdf}
\usepackage{amsmath,amssymb}
\usepackage{float}
   \usepackage[all]{xy}
\usepackage{subfigure}
\usepackage{xcolor}
\usepackage{soul} 
\usepackage{blkarray}

\begin{document}

\title{Evolution of cooperation in deme-structured populations on graphs}
\author{Alix Moawad}
\altaffiliation[Present address: ]{Department of Biosystems Science and Engineering (D-BSSE), ETH Zurich, CH-4058 Basel, Switzerland}
  \affiliation{Institute of Bioengineering, School of Life Sciences, École Polytechnique Fédérale de Lausanne (EPFL), CH-1015 Lausanne, Switzerland}
  \affiliation{SIB Swiss Institute of Bioinformatics, CH-1015 Lausanne, Switzerland} 
 
 \author{Alia Abbara}
  \affiliation{Institute of Bioengineering, School of Life Sciences, École Polytechnique Fédérale de Lausanne (EPFL), CH-1015 Lausanne, Switzerland}
  \affiliation{SIB Swiss Institute of Bioinformatics, CH-1015 Lausanne, Switzerland}
 
  \author{Anne-Florence Bitbol}
  \email[Corresponding author: ]{anne-florence.bitbol@epfl.ch}
  \affiliation{Institute of Bioengineering, School of Life Sciences, École Polytechnique Fédérale de Lausanne (EPFL), CH-1015 Lausanne, Switzerland}
  \affiliation{SIB Swiss Institute of Bioinformatics, CH-1015 Lausanne, Switzerland}

\begin{abstract}
	Understanding how cooperation can evolve in populations despite its cost to individual cooperators is an important challenge. Models of spatially structured populations with one individual per node of a graph have shown that cooperation, modeled via the prisoner's dilemma, can be favored by natural selection. These results depend on microscopic update rules, which determine how birth, death and migration on the graph are coupled. Recently, we developed coarse-grained models of spatially structured populations on graphs, where each node comprises a well-mixed deme, and where migration is independent from division and death, thus bypassing the need for update rules. Here, we study the evolution of cooperation in these models in the rare migration regime, within the prisoner's dilemma. We find that cooperation is not favored by natural selection in these coarse-grained models on graphs where overall deme fitness does not directly impact migration from a deme. This is due to a separation of scales, whereby cooperation occurs at a local level within demes, while spatial structure matters between demes.
\end{abstract}

\maketitle

\section{Introduction}
Cooperation between individuals is commonly observed in living systems, at the cell scale~\cite{Hunter91,axelrod81,celiker2013cellular, mitchison2004t}, between animals~\cite{dugatkin1997cooperation, raihani2012punishment,chase1980cooperative, dugatkin2000cheating}, and in human societies~\cite{johnson2003puzzle, rand2013human, puurtinen2009between}. For instance, bacteria often produce diffusible resources (e.g.\ enzymes, toxins or signaling molecules) that can benefit neighboring bacteria, including those that do not produce this resources (cheaters)~\cite{Borenstein13,Cremer19}. Because producing these resources is generally costly, natural selection \textit{a priori} works against the producers of public goods, and favors cheaters. Therefore, understanding how cooperation can evolve and spread is an important question~\cite{Wingreen06,Cremer19}, which has led to intense debates. Kin selection can lead to the evolution of cooperation under some conditions~\cite{Hamilton64a,Hamilton64b}: if the individuals that benefit from cooperation are related to cooperators, then cooperation can be selected even if it carries a cost. Spatial structure can also favor cooperation in some models~\cite{ohtsuki2006evolutionary,ohtsuki2006simple,chuang2009simpson,Melbinger10,Cremer12,Cremer19}. Since spatial structure generally entails that closely located individuals are genetically related, its ability to promote cooperation is related to kin selection~\cite{Kay20}. In evolutionary game theory, individuals play games such as the prisoner's dilemma~\cite{doebeli2005,Turner99} or the snowdrift game~\cite{doebeli2005,Gore09} using fixed strategies. Their fitnesses depend on the composition of the surrounding population, via the payoff they receive from the games played with their neighbors~\cite{sigmund1999evolutionary, weibull1997evolutionary,traulsen2009stochastic, hummert2014evolutionary}. This allows to model cooperation in a straightforward way and to study its evolution. In these models, each individual is located on a node of a graph~\cite{lieberman2005evolutionary}, allowing to investigate the impact of complex spatial structures on the evolution of cooperation~\cite{ohtsuki2006simple, ohtsuki2006evolutionary}. It was found that in the prisoner's dilemma game, a cooperator mutant can be favored by natural selection in some graphs under specific update rules that prescribe how individuals are chosen to divide, die, and replace each other~\cite{ohtsuki2006simple, ohtsuki2006evolutionary}.

Recently, we developed more general models of spatially structured populations on graphs, where each node comprises a well-mixed deme instead of a single individual, and where migration between demes is independent from division and death~\cite{marrec2021toward,Abbara23}. These models therefore do not require an update rule, contrary to other models with demes on graphs~\cite{Houchmandzadeh11,yagoobi2021fixation, yagoobi2023}. We showed that for mutants with constant (non-frequency-dependent) fitness, some spatial structures with specific migration asymmetries can amplify natural selection in the regime of rare migration~\cite{marrec2021toward}. We further found that suppression of selection is pervasive with more frequent migrations, in the branching process regime~\cite{Abbara23}. Importantly, these results do not depend on update rules, contrary to models with one individual per node where suppression or amplification of natural selection for a given graph strongly depends on the update rule choice~\cite{Kaveh15, Pattni15, Hindersin15, tkadlec2020limits}. In the rare migration regime, this dependence on update rules was replaced by a dependence on migration asymmetry, which can be experimentally tuned and measured, while update rules model assumptions about the microscopic dynamics at the cell scale, which are generally not strictly followed by natural or experimental populations.

Can spatial structure favor the evolution of cooperation in these deme-structured models, as in models with one individual per node of the graph? To address these questions, we consider the prisoner's dilemma in spatially structured populations within the frameworks of~\cite{marrec2021toward,Abbara23}. In these models, the fate of mutants is impacted by the choice of graph structure and specific migration rates between demes, but the average fitness of a deme does not directly impact the intensity of its outgoing (or incoming) migrations. In other words, selection is essentially soft~\cite{Wallace75}, as we aimed for a minimal incorporation of spatial structure. Note however that cooperation can be favored via a coupling between composition and population size~\cite{chuang2009simpson,Melbinger10,Cremer12,Cremer19}. 

We first lay out our models of deme-structured populations on graphs with cooperation in the case of the prisoner's dilemma. Next, we consider a single cooperator (mutant), placed in a population entirely composed of defectors (wild-types), and we ask what the probability that it takes over (i.e.\ fixes) is. We address this question in different graphs, for rare migrations. We compare the fixation probability of a single cooperator to that of a single defector, and to that of a neutral mutant, to determine whether cooperation is favored or not. We also evaluate the impact of structure by comparing the fixation probability of a cooperator mutant in a structure to that in a well-mixed population with the same total size. We find that cooperators are not favored by natural selection in our models. The impact of spatial structure on their fixation probability reduces to that observed for mutants with fixed fitness disadvantage. This is due to a separation of scales, with cooperation occurring within demes, while spatial structure matters between demes. We make complete comparisons with models with one individual per node of the graph, including formal calculations in the Appendix, where we show the points that differ between these two classes of models.

\section{Model and methods}

\subsection{Spatially structured populations with demes on graphs}

\begin{figure}[htbp]
    \centering
    \includegraphics[width=6cm]{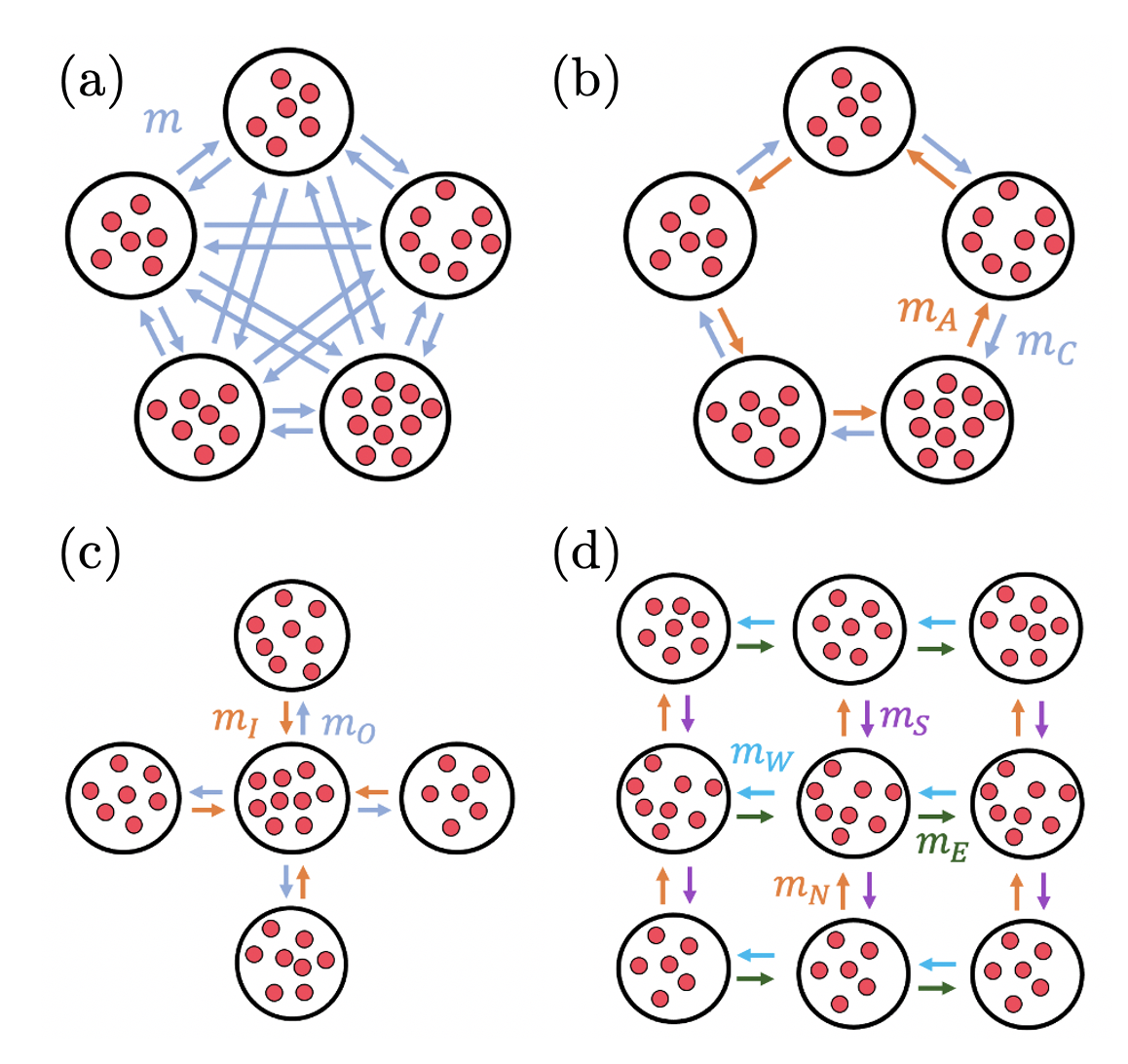} 
    \caption{Highly symmetric population structures: (a) clique, (b) cycle, (c) star and (d) $3\times 3$ regular lattice. (a), (b), and (c) have 5 demes, while (d) has 9 demes, represented by large white circles. Each deme holds a well-mixed population of microorganisms represented by red circles. Individuals can migrate from one deme to another along the depicted arrows, following the written migration rates. }
    \label{structures}
\end{figure}

We model spatially structured populations on directed graphs. Each of the $D$ nodes of the graph contains a well-mixed subpopulation or deme. The specific graphs considered here are shown in Fig.~\ref{structures}. Two types of individuals make up the population: mutants ($M$) and wild-types ($W$). We compare two different models, which differ by how deme size is regulated. In the first one, each individual can divide, die and migrate with given rates, deme sizes are limited by a carrying capacity, and we consider the regime where they fluctuate around a steady-state value \cite{marrec2021toward}. In the second one, inspired by serial transfer experiments, the population undergoes an alternation of growth and bottleneck phases. Starting from a bottleneck state, each deme undergoes exponential growth, and then some individuals are sampled and migrate to form a new bottleneck state \cite{Abbara23}. Formally, the first model is close to a Moran model, but deme size is not strictly fixed \cite{marrec2021toward}. Meanwhile, the second one is close to a Wright-Fisher model \cite{Abbara23}. We mainly study the first one, which is more directly comparable to models with one individual per node~\cite{lieberman2005evolutionary,ohtsuki2006evolutionary,ohtsuki2006simple}. The second model allows us to extend our results to more frequent migrations.

Within our models of spatial structure, we consider an evolutionary game theory setup that provides a simple description of cooperation, following \cite{ohtsuki2006evolutionary,ohtsuki2006simple,traulsen2009stochastic}. Our two types of individuals then correspond to cooperators and defectors,
interacting within a prisoner's dilemma with payoff matrix 
\begin{equation}\label{payoffmatrix}
P = 
\begin{pmatrix}
  b-c  & -c\\
  b & 0 \\
\end{pmatrix}\,,
\end{equation} 
where $b > c > 0$. If a cooperator interacts with another cooperator, then both receive a benefit $b$ but also lose a cost $c$; this yields $b-c$ in the upper left corner of $P$. If a cooperator interacts with a defector, then the cooperator does not receive anything, but it loses $c$ because of the interaction (upper right corner), while the defector receives $b$ (bottom left corner). Finally, if two defectors interact, both get 0. Note that here, being a cooperator or a defector in our model is an intrinsic (genetic) characteristic of an individual. Individual fitnesses are expressed as $f = 1-w+w\pi$ where $w$ is the intensity of selection and $\pi$ the total payoff received by the individual upon interactions with all its partners. We will usually consider cooperators as mutants ($M$) and defectors as wild-types ($W$), and ask how likely it is that cooperator mutants take over.

We consider that interactions between individuals occur inside demes, which are well-mixed. Thus, each individual interacts with all other individuals in the same deme. Denoting by $k$ the number of mutants in a deme of size $N$, the payoff $\pi_M(k,N)$ for a mutant individual and $\pi_W(k,N)$ for a wild-type individual in this deme read~\cite{traulsen2009stochastic}:
\begin{equation}
\begin{cases}\label{payoffs}
        \pi_M(k,N)&=  (b-c)\frac{k-1}{N-1} - c \frac{N-k}{N-1} \,, \\
        \pi_W(k,N)&= b\frac{k}{N-1}\,.
\end{cases}
\end{equation}
In these expressions, $(k-1)/(N-1)$ is the fraction of mutants among the individuals that a mutant of focus can interact with (i.e. all but itself), while $(N-k)/(N-1)$ is the fraction of wild-types among them. Similarly, $k/(N-1)$ is the fraction of mutants among the individuals that a wild-type of focus can interact with. In each of our two models, fitness can be expressed from these payoffs, giving rise to frequency-dependent fitnesses (see below).

\subsection{Model with carrying capacities}

\begin{figure}[htbp]
    \centering
    \includegraphics[width=6cm]{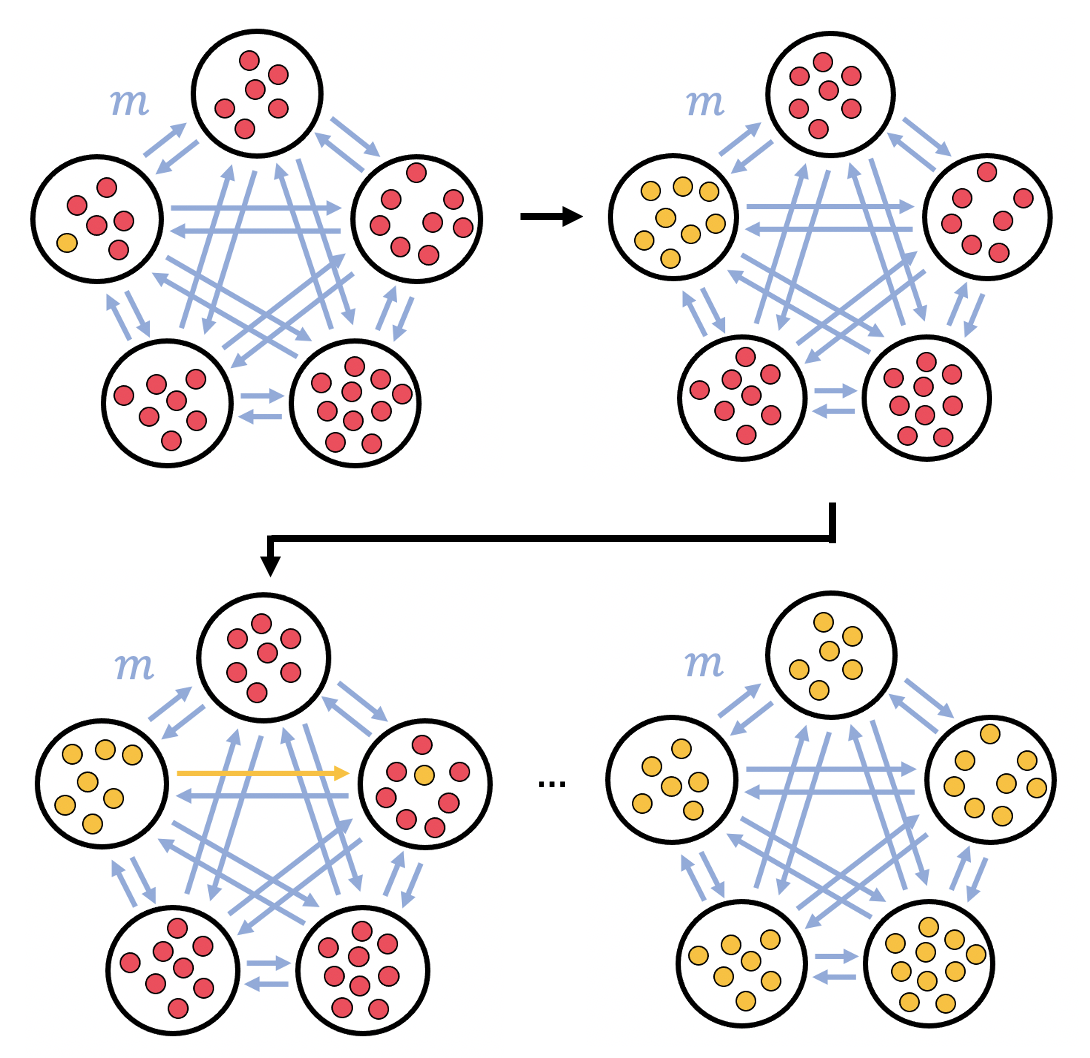} 
    \caption{Fixation of a mutant in a clique. We start with a single mutant individual (yellow) in one deme, while all the other individuals are wild-types (red). In the rare migration regime, mutants first fix in the initial deme before one migration event to another deme occurs. This process is repeated until mutants take over the complete structure.}
    \label{Evolutionprocess}
\end{figure}

We first consider the model introduced in \cite{marrec2021toward}. In this model, each deme has maximum carrying capacity $K$. Besides, each individual has fitness $f_M$ or $f_W$, and death rate $g_M=g_W=g$, as we focus on selection on division. Here, fitness represents the maximal division rate of microorganisms, reached in exponential growth. The division rate of individuals of type $A=M,W$ in deme $i$ is given by the logistic function $f_A(1 - N_i/K)$, where $N_i$ is the number of individuals in deme $i$. As in \cite{marrec2021toward}, we focus on the rare migration regime, where fixation or extinction within a deme occurs on much shorter time scales than migrations (see Fig. \ref{Evolutionprocess}). In this regime, we describe the evolution of the microbial population as a Markov process where each event is a migration and where each deme is either fully mutant or fully wild-type~\cite{Slatkin81,marrec2021toward}. Note that we will present an extension to more frequent migrations using our second model (see below). We focus on the regime where deme sizes $N_i$ fluctuate weakly around their deterministic steady-state value $K(1-g/f_W)$ (where we have neglected differences between $f_M$ and $f_W$, which is acceptable for weak selection, and where we have neglected migrations, which is acceptable if they are rare). Note that the deterministic steady-state size is equal to the maximum carrying capacity $K$ if $g=0$. In what follows, we will refer to $K$ as the deme carrying capacity, for simplicity. We approximate the fixation process of a type within a deme by a Moran process \cite{moran1958random, ewens2004mathematical} for a population of size $K(1-g/f_W)$, thus neglecting size fluctuations. Here, we implement logistic regulation in the division rate, which decreases if $N_i$ grows. Appendix~\ref{OtherModel} briefly discusses a variant where it is instead implemented in the death rate. Our conclusions are not affected. 

Denoting by $k$ the number of mutants in a deme of size $N$ as above, the frequency-dependent fitness $f_M(k,N)$ for a mutant individual and $f_W(k,N)$ for a wild-type individual in this deme read:
\begin{equation}
    \begin{cases}
        f_M(k,N) =1-w+w\pi_M(k,N)\,,\\
        f_W(k,N) =1-w+w\pi_W(k,N)\,,\label{fit}
    \end{cases}
\end{equation}
where payoffs $\pi_M$ and $\pi_W$ are given by Eq.~(\ref{payoffs}). Under this convention, where fitnesses depend linearly on payoffs, we will focus on the weak selection regime defined by $w\ll 1$ and $wN\ll 1$, where both $f_M$ and $f_W$ are close to the base fitness $f=1$~\cite{ohtsuki2006simple,ohtsuki2006evolutionary}. We will then briefly discuss an alternative convention where fitnesses depend exponentially on payoffs~\cite{Traulsen08}, which allows an extension of our results beyond the weak selection regime.

We perform analytical calculations and stochastic simulations. For the latter, we use a Gillespie algorithm \cite{gillespie1976general, gillespie1977exact,marrec2021toward}, described in detail in Appendix~\ref{App_simu}. We will refer to this model in figures as the ``carrying-capacity'' (C. C.) one.

\subsection{Model with serial dilutions}
\label{dilution_model}
Recently, in Ref.~\cite{Abbara23}, we introduced a new model of spatially structured populations on graphs, directly inspired by the batch culture setups with serial transfers that are used in many evolution experiments~\cite{Lenski91,Elena03,Good17,Kryazhimskiy12,Nahum15,France19,Chen20,Kassen} and are important in ecology~\cite{Erez20}. This \emph{serial dilution model} is a variant of the Wright-Fisher model, and features demes placed on the nodes of a graph with possible migrations along its edges, like the carrying-capacity model defined above. Each deme holds a well-mixed subpopulation, and within-deme interactions implement cooperation between individuals following the payoff matrix (\ref{payoffmatrix}). In this model, all demes have the same bottleneck size $B$, and repeatedly undergo a two-step process. To clarify notations, quantities common to the two models will be denoted with a prime for the serial dilution model.
\begin{enumerate}
    \item First, subpopulations grow exponentially in each deme during a fixed growth time $t$. Time within this growth phase is denoted by $\tau \in [0,t]$. Starting at $\tau=0$ from $k'(0)$ mutants and $l'(0)$ wild-types in a deme at bottleneck, such that $k'(0)+l'(0)=N'(0)=B$, these numbers grow as:
\begin{align}
\label{growth_phase}
\begin{split}
    \dfrac{dk'}{d\tau}(\tau) &= f'_M(\tau) k'(\tau) \,, \\
    \dfrac{dl'}{d\tau}(\tau) &= f'_W(\tau) l'(\tau) \,,
\end{split}
\end{align}
with rates $f_M(\tau)$ and $f_W(\tau)$ reading
\begin{equation}
    \begin{cases}
        f'_M(\tau) =1 - w' + w'\pi_M (\tau)\,,\\
        f'_W(\tau) =1 - w' + w' \pi_W (\tau)\,.\label{fit_dilution}
    \end{cases}
\end{equation}
Here $\pi_M (\tau)$ is a shorthand for the payoff $\pi_M (k'(\tau),N'(\tau))$ given in Eq.~(\ref{payoffs}), with population size $N'(\tau)=k'(\tau)+l'(\tau)$ [and similarly for $\pi_W (\tau)$].
The growth rates in Eq.~(\ref{fit_dilution}) are chosen to have a similar form to the fitnesses given in Eq.~(\ref{fit}). Here too, the payoffs obtained from the prisoner's dilemma within demes directly impact the ability of cells to divide. 
\item Then, a dilution and migration step occurs (see details in Section~\ref{dilution_details} of the Appendix), bringing the system to a new bottleneck state. For each deme, we perform a multinomial sampling to pick exactly $B$ individuals that form the new bottleneck state of this deme. These incoming individuals are sampled from all demes, according to migration probabilities (all incoming migration probabilities to a given deme sum to 1).
\end{enumerate}
In addition to bridging with experiments, this model has the advantage of facilitating the treatment of frequent migrations. Indeed, in our carrying capacity model under frequent asymmetric migrations, some demes can have a steady-state size that exceeds $K$, which prevents division there. Choosing a division rate independent from $N$ and a death rate proportional to $N/K$ would instead result in substantially faster turnover in these demes. These situations lack realism and strongly depend on the exact form of deme size regulation. The serial dilution model does not suffer from this drawback.

We perform stochastic simulations under this serial dilution model~\cite{Abbara23}, and we refer to it in figures as the ``dilution'' one. Considering the serial dilution model allows us to test the robustness of our conclusions from the carrying-capacity model in the rare migration regime, and to extend our study beyond this regime.

To quantitatively compare results from the two models, we choose the bottleneck size $B$ in the dilution model to be equal to the steady-state size of the demes $K(1-g/f_W)$ in the carrying-capacity model. Furthermore, in the dilution model, we introduce the effective intensity of selection $2w't$. Indeed, it plays the same role as the intensity of selection $w$ in the carrying-capacity model, see Section~\ref{map_models} of the Appendix (in particular, the factor of 2 comes from the difference between a Moran and a Wright-Fisher description in the diffusion limit).

\section{Results}

\subsection{Fixation probability in a deme and in a structured population with rare migrations}
\label{subsec:res1}

In the rare migration regime, fixation of a mutant in the whole population starts by fixation in the deme where the mutant was introduced (see Methods and Fig.~\ref{Evolutionprocess}). Thus, we can write the fixation probability $\rho_M^\textrm{struct}$ of a mutant in the structure as 
\begin{equation} \label{pfix_gal}
    \rho_M^\textrm{struct}=\rho_M\Phi_1\,,
\end{equation}
where $\rho_M$ is the fixation probability of one mutant in a well-mixed deme where all other individuals are wild-type, while $\Phi_1$ is the fixation probability of the mutant type in the whole population starting from one fully mutant deme and $D-1$ wild-type demes. 

Let us first focus on $\rho_M$, and let us compute it in the carrying-capacity model. Neglecting deme size fluctuations in the fixation process within a deme (see Methods) and denoting by $N$ the number of individuals in the deme, we can follow the calculation of \cite{traulsen2009stochastic}, which yields:
\begin{equation}\label{rhoM}
    \rho_M = \left(1+\sum_{j=1}^{N-1} \prod_{k=1}^{j} \dfrac{f_W(k,N)}{f_M(k,N)} \right)^{-1}\,.
\end{equation}  
Note that $N$ corresponds to the deterministic steady-state deme size $N=K(1-g)$, where we have neglected the impact of fitness differences on $N$ and used $f_W\approx f_M\approx 1$, as we focus on the weak-selection regime (this amounts to neglecting terms of order $wg/(f-g)$, which are explicitly written in~\cite{marrec2021toward} -- this is acceptable if $g\ll f$, which is realistic).
Similarly, the fixation probability $\rho_W$ of one wild-type in a well-mixed deme where all other individuals are mutant reads
\begin{equation}\label{rhoW}
    \rho_W = \left(1+\sum_{j=1}^{N-1} \prod_{k=1}^{j} \dfrac{ f_M(N-k,N)}{ f_W(N-k,N)}\right)^{-1}\,.
\end{equation}
Inserting the fitness expressions from Eq.~(\ref{fit}) into Eqs.~(\ref{rhoM}) and (\ref{rhoW}), and performing an expansion in $w$ in the weak selection regime $w\ll 1$ and $wN\ll 1$ yields:
\begin{align} \label{rhoMa}
\rho_M = \frac{1}{N}\left\{ 1-\frac{w}{2}\left[b+c(N-1)\right]\right\} + O(w^2)\,,\\
\rho_W = \frac{1}{N}\left\{ 1+\frac{w}{2}\left[b+c(N-1)\right]\right\} + O(w^2)\,. \label{rhoWa}
\end{align}
We note that $\rho_M < 1/N < \rho_W$: we recover the result that in a well-mixed population, natural selection favours defectors over cooperators.

To calculate the fixation probability $\rho_M^\textrm{struct}$ of a mutant in the whole structured population, we need to express the probability $\Phi_1$ that the mutant type fixes, starting from one fully mutant deme, see Eq.~(\ref{pfix_gal}). Importantly, $\Phi_1$ depends on the specific graph that is considered for the structured population. Below, we focus on the simple graphs shown in Fig.~\ref{structures}, and investigate how each of them impacts the evolution of cooperation. To compute $\Phi_1$, we notice that in the rare migration regime, the state of the population can be fully described by specifying which demes are fully mutant and which ones are fully wild-type. Furthermore, this state evolves due to migration events, which may change it if they result into fixation in the destination deme. The evolution of the state of the population is a Markov chain~\cite{Slatkin81,marrec2021toward}. 

\subsection{Evolution of cooperation in the clique and cycle}
\label{circs}

Let us first consider the clique and the cycle, where all demes are equivalent. To compute $\Phi_1$, and more generally the fixation probability $\Phi_i$ when starting from $i$ fully mutant demes (consecutive in the case of the cycle) and $D-i$ fully wild-type demes, we follow the method we used in~\cite{marrec2021toward}, which is based on~\cite{traulsen2009stochastic}. Briefly, we write down a recurrence relation on $\Phi_i$ by discriminating over all possible outcomes of the first migration event: either $i$ increases by one if a mutant migrates to a wild-type deme and fixes there, or it decreases by one in the opposite case, or it stays constant in all other cases (see Appendix \ref{app:cycle} and~\cite{marrec2021toward} for details). Solving this recurrence relation yields
\begin{equation}\label{eq_clique_cycle}
    \Phi_1^{\textnormal{clique}}=\Phi_1^{\textnormal{cycle}} = \frac{1-\gamma}{1-\gamma^D}\,,
\end{equation}
with 
\begin{equation}\label{gamma}
    \gamma = \frac{\rho_W}{\rho_M}\,.
\end{equation}
The reasoning made with frequency-independent fitness in~\cite{marrec2021toward} thus extends to the present case with cooperation, and accordingly, the formal expression of $\Phi_1$ is the same in both cases. However, the expression of $\gamma$ differs because the fixation probabilities within a deme are impacted by the cooperation model. In the present case, $\gamma$ can be expressed using Eqs.~(\ref{rhoMa}) and~(\ref{rhoWa}). Note that Eq.~(\ref{eq_clique_cycle}) is independent from migration rates. In particular, in the case of the cycle, Eq.~(\ref{eq_clique_cycle}) does not depend on the ratio $m_A/m_C$ of the anticlockwise migration rate $m_A$ to the clockwise one $m_C$, see Fig.~\ref{structures}. 

In~\cite{marrec2021toward}, we further showed that all circulation graphs (such that the sum of incoming migration rates to any given deme is equal to the sum of outgoing ones from that deme), including the cycle, have the same fixation probability, thereby extending the circulation theorem which holds for graphs with one individual per node~\cite{lieberman2005evolutionary}. The circulation theorem further extends to the present case with cooperation. Indeed, the exact same proof as in~\cite{marrec2021toward} holds, albeit with a different $\gamma$. Therefore, all circulation graphs have the same fixation probability of cooperator mutants, given by Eqs.~(\ref{eq_clique_cycle}) and~(\ref{gamma}). We will henceforth denote it by $\Phi_1^{\textnormal{circ.}}$.

Using Eqs.~(\ref{rhoMa}) and~(\ref{rhoWa}) to express Eq.~(\ref{eq_clique_cycle}) yields
\begin{equation}\label{eq_clique_cycle_approx}
    \Phi_1^{\textnormal{circ.}}= \frac{1}{D}\left\{1-\frac{D-1}{2}[b+c(N-1)]w+O(w^2)\right\}\,,
\end{equation}
which gives the fixation probability of a mutant in the whole population, using Eq.~(\ref{pfix_gal}) and Eq.~(\ref{rhoMa}):
\begin{equation}\label{phiM_clique_cycle_approx}
    \rho_M^{\textnormal{circ.}}= \frac{1}{ND}-\frac{1}{2N}[b+c(N-1)]w+O(w^2)\,.
\end{equation}
Importantly, this fixation probability is smaller than the neutral value $1/(ND)$ for all $b>c>0$: cooperation is never favored by natural selection in circulation graphs, including the clique or the cycle. It is also very similar to the fixation probability of a mutant in a well-mixed population of size $ND$ [obtained from Eq.~(\ref{rhoMa}) by replacing $N$ with $ND$] -- both coincide for $N\gg 1$ (see also~\cite{marrec2021toward}).

\begin{figure}[htbp]
    \centering
    \includegraphics[angle=90,origin=c,width=\columnwidth]{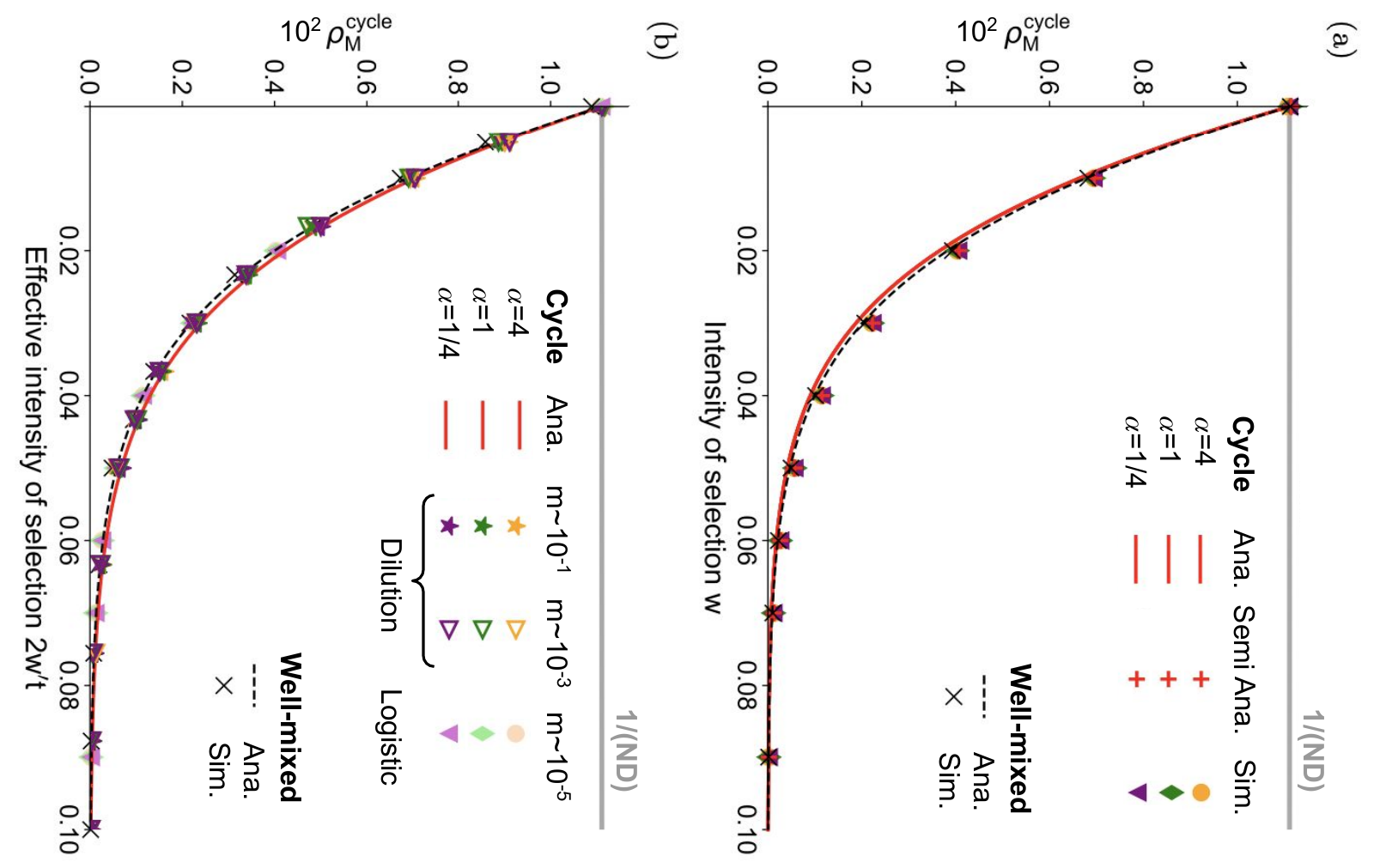} 
    \caption{(a) Mutant fixation probability $\rho^{\textnormal{cycle}}_{M}$ in the cycle versus intensity of selection $w$ for different migration asymmetries $\alpha=m_A/m_C$, in the rare migration regime. Solid lines represent analytical predictions; red ``+" markers are semi-analytical predictions; other markers are simulations. The well-mixed population case (black dashed line), and the neutral mutant case (gray line) are shown for reference. Analytical and semi-analytical predictions are obtained using Eqs.~(\ref{pfix_gal},\ref{eq_clique_cycle}), with $\rho_M$ and $\rho_W$ calculated either using Eqs.~(\ref{rhoM},\ref{rhoW}) that hold for fixed deme size, or using simulations on well-mixed demes ($10^7$ replicates) to avoid errors from assuming fixed deme size. Pure simulation results are obtained over $10^6$ replicates with ($m_A$, $m_C$) $\times 10^{5}$  = (4,1), (1,1) and (1,4).  
    \\ (b) Mutant fixation probability $\rho^{\textnormal{cycle}}_{M}$ in the cycle versus effective intensity of selection $2w't$ in the dilution model, for different migration asymmetries and intensities. The solid red line represents the analytical prediction in the rare migration regime in the dilution model, from Eqs.~(\ref{pfix_gal},\ref{eq_clique_cycle}), using Eqs.~(\ref{rhoM_dil}, \ref{rhoW_dil}) for $\rho_M$ and $\rho_W$. The well-mixed population case (black dashed line) in the dilution model and simulations for the carrying-capacity (C. C.) model in the rare migration regime (versus $w$) are shown for comparison. Dilution simulation results are obtained over $10^6$ realizations for ($m_A$, $m_C$) $\times 10$ and $\times 10^{3}$ = (4,1), (1,1) and (1,4). \\
    In this figure, we consider $D = 5$ demes with carrying capacity $K = 20$, death rate $g=0.1$ yielding steady-state size $N=18$, and benefit and cost $b = 2$ and $c = 1$. The well-mixed population holds $D N =90$ individuals. }
    \label{rhoMcycle}
\end{figure}

In Fig.~\ref{rhoMcycle}(a), we compare numerical simulation results and analytical predictions for the fixation probability $\rho_M^{\textnormal{cycle}}$ in the cycle. 
We find that stochastic simulation results do not depend on migration asymmetry $\alpha=m_A/m_C$, consistently with our analytical calculations of $\rho_M^{\textnormal{cycle}} = \rho_M \cdot \Phi_1^{\textnormal{cycle}}$. A slight discrepancy can however be observed between analytical predictions and simulation results. It arises from the constant-size approximation we made to compute the analytical expressions of $\rho_M$ and $\rho_W$, which are both involved in the expression of $\rho_M^{\textnormal{cycle}}$. Indeed, using Gillespie simulation results obtained on well-mixed demes for $\rho_M$ and $\rho_W$ instead of their analytical expressions for fixed size within the expression $\rho_M^{\textnormal{cycle}} = \rho_M \cdot \Phi_1^{\textnormal{cycle}}$ yields an excellent agreement with simulation results [see Fig.~\ref{rhoMcycle}(a)]. 

In Fig.~\ref{rhoMcycle}(b), we explore more frequent migrations, using the serial dilution model introduced above. Our simulation results suggest that the fixation probability for a cycle remains the same for more frequent migrations.
Besides, as expected, matching the effective intensity of selection $2w't$ in this model to $w$ in our first carrying-capacity model, simulations from both models give close results for rare migrations. The rare migration analytical prediction for the dilution model is obtained with Eq.~(\ref{eq_clique_cycle}), but replacing $\rho_M$ and $\rho_W$ with their expressions $\rho'_M$ and $\rho'_W$ derived in the dilution model (see details in Section~\ref{dilution_details} of the Appendix). Note that~\cite{Abbara23} extended the circulation theorem to the dilution model without cooperation, both for rare migrations, and for frequent migrations within the branching process approximation.

Our results stand in contrast to those of~\cite{ohtsuki2006simple}, where the present model of cooperation was studied in a cycle graph with one individual per node. There, it was found that cooperators could be favored by natural selection under the death-Birth (dB) and imitation update rules, but not under the Birth-death (Bd) update rule (the uppercase B indicates that selection on fitness happens at the birth step). Specifically, natural selection was found to favor cooperation in the dB case if $b/c>2+4/(D-4)$~\cite{ohtsuki2006simple}. To understand the formal origin of this difference, we revisit the derivation of~\cite{ohtsuki2006simple} in Appendix \ref{app:cycle}. The key difference is that in our model, cooperative interactions occur within a deme and are not involved in migration events. Conversely, in the model of ~\cite{ohtsuki2006simple}, they impact the spread of mutants on the graph since birth, death and migration events are coupled via update rules.

\newpage

\subsection{Evolution of cooperation in the star}

Let us now consider the star graph, where the center sends out individuals to any leaf with rate $m_O$, and receives individuals from any leaf with rate $m_I$ [see Fig.~\ref{structures}(c)]. This graph structure has been extensively studied for mutants with constant fitness. In evolutionary graph theory with one individual per node, the star amplifies or suppresses natural selection depending on the update rule~\cite{lieberman2005evolutionary,Kaveh15,Hindersin15,Pattni15}. For instance, it is an amplifier of selection using the Bd update rule, but a suppressor with the dB rule. In evolutionary graph theory with nodes containing subpopulations, the evolutionary outcome also depends on the specific update mechanism~\cite{yagoobi2023}.
In our model with rare migrations~\cite{marrec2021toward} that does not rely on a specific update rule, the star's behavior depends on migration asymmetry $\alpha=m_I/m_O$, as $\alpha>1$ amplifies selection while $\alpha<1$ suppresses it. This same outcome holds for the rare migration regime in the serial dilution model~\cite{Abbara23}. However, for more frequent migrations, the star becomes a suppressor for all migration asymmetries in the branching process regime~\cite{Abbara23}. How does the star graph impact the fixation of a cooperator, in the rare migration regime and beyond? 

To address this question, we extend the calculation of~\cite{marrec2021toward} to our cooperation model, and we express the fixation probability $\Phi^{\textnormal{star}}_1$ starting from one fully mutant deme, uniformly chosen among all demes, in the star graph. As in~\cite{marrec2021toward}, we obtain
\begin{equation}\label{phi1starfrom01and10}
    \Phi^{\textnormal{star}}_1 = \frac{(1-\gamma^2)[\gamma + \alpha D + \gamma \alpha^2 (D-1)]}{D(\alpha+\gamma)[1+\alpha \gamma - \gamma^D (\alpha + \gamma)^{2-D} (1+\alpha \gamma)^{D-1}]}\,,
\end{equation}
where $\gamma$ is given by Eq.~(\ref{gamma}), while $\alpha = m_I/m_O$ is the ratio of the migration rate incoming to the center from a leaf $m_I$, to that outgoing from the center to a leaf $m_O$, see Fig. \ref{structures}. Eq.~(\ref{phi1starfrom01and10}) shows that migration asymmetry $\alpha$ directly impacts the fixation probability in the star. In~\cite{marrec2021toward}, we demonstrated that the star amplifies natural selection for $\alpha>1$ and suppresses it for $\alpha<1$. For $\alpha=1$, the fixation probability reduces to Eq.~(\ref{eq_clique_cycle}), as the star is then a circulation graph.

\begin{figure}[htbp]
    \centering
    \includegraphics[angle=90,origin=c,width=\columnwidth]{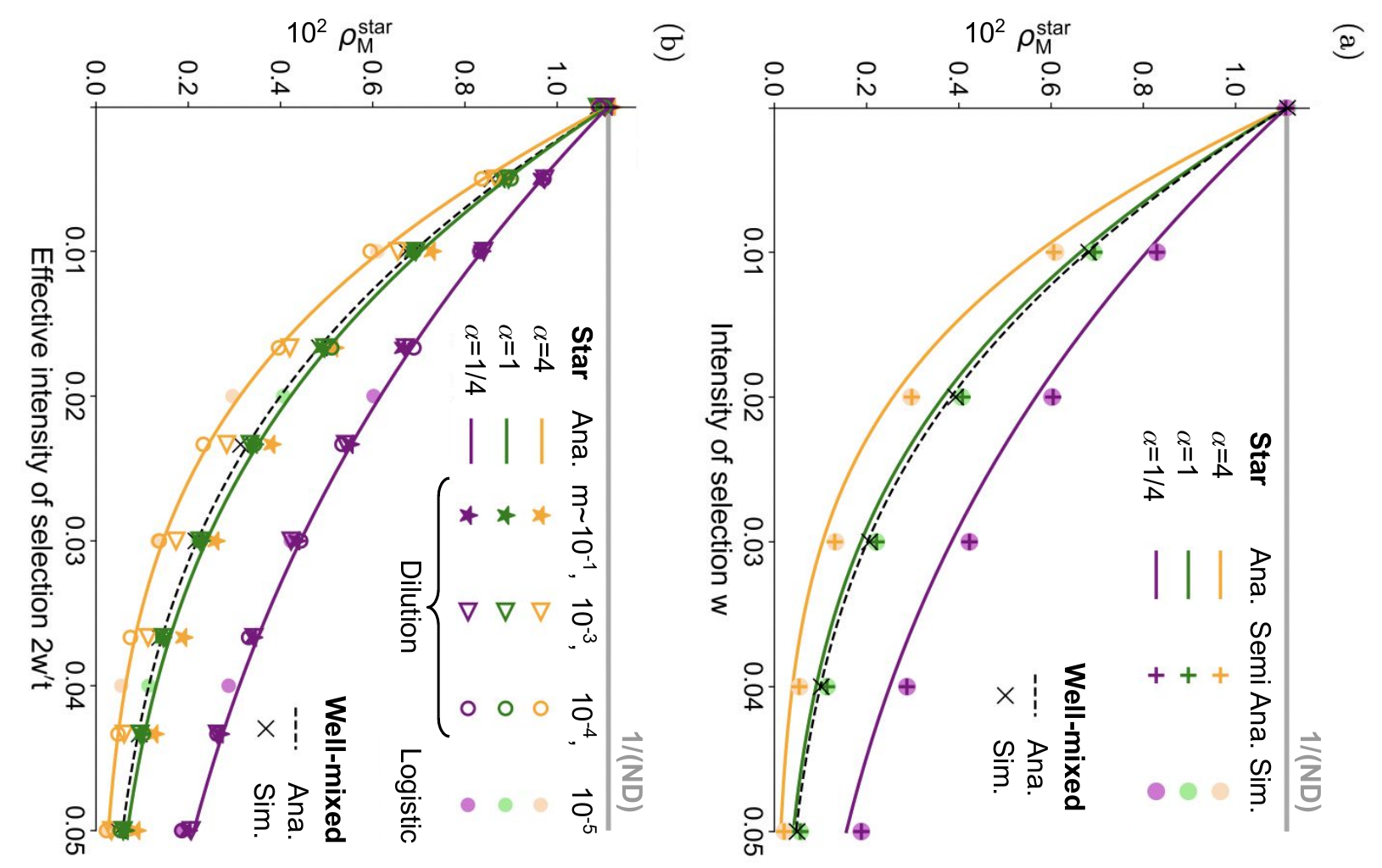}
    \caption{(a) Mutant fixation probability $\rho^{\textnormal{star}}_{M}$ in the star versus intensity of selection $w$ for different values of migration asymmetry $\alpha=m_I/m_O$, in the rare migration regime. Pure simulation results are obtained over $10^6$ replicates, with ($m_I$, $m_O$) $\times 10^{5}$  = (4,1), (1,1) and (1,4). 
    \\ (b) Mutant fixation probability $\rho^{\textnormal{star}}_{M}$ in the star for the dilution model versus effective intensity of selection $2w't$, for different migration asymmetries and intensities. Simulations for the dilution model are obtained over $10^6$ realizations for ($m_I$, $m_0$)$\times 10$ and $\times 10^3$ = (4,1), (1,1) and (1,4). \\
    In both panels, analytical predictions in the rare migration regime use Eq.~(\ref{phi1starfrom01and10}) instead of Eq.~(\ref{eq_clique_cycle}). They yield values close to the well-mixed case for $\alpha$=1 (green [middle] solid line), larger values for $\alpha$=1/4 (purple [upper] solid line), and smaller values for $\alpha$=4 (yellow [bottom] solid line).}
    \label{rhoMstar}
\end{figure}

Using Eqs.~(\ref{rhoMa}) and~(\ref{rhoWa}) to express Eq.~(\ref{phi1starfrom01and10}) yields
\begin{align}\label{eq_star_approx}
    \Phi_1^{\textnormal{star}}&= \frac{1}{D}\bigg\{1-\frac{\alpha(D-1)[\alpha(D-2)+2]}{(\alpha+1)[\alpha(D-1)+1]}\nonumber\\
    &\times[b+c(N-1)]w+O(w^2)\bigg\}\,.
\end{align}
Studying the ratio between Eq.~(\ref{eq_star_approx}) and Eq.~(\ref{eq_clique_cycle_approx}) shows that the fixation probability of a cooperator mutant is larger in the star than in the clique for $\alpha<1$ (i.e. when the star behaves as a suppressor of selection~\cite{marrec2021toward}) and smaller for $\alpha>1$ (i.e. when the star behaves as an amplifier of selection~\cite{marrec2021toward}). This result is confirmed by stochastic simulations in Fig.~\ref{rhoMstar}(a). These effects are purely due to the population structure and not to the cooperation model. In other words, what we find here holds as well for a deleterious mutant without frequency-dependent fitness (see~\cite{marrec2021toward}). Simulations in Fig.~\ref{rhoMstar}(b) show that cooperation does not impact the star's behavior in the dilution model either, compared to the case of mutants with constant fitness~\cite{Abbara23}. Indeed, the rare migration regime recovers results from the carrying-capacity model, but as migrations become frequent, the star becomes a suppressor of natural selection regardless of migration asymmetry~\cite{Abbara23}.
\vspace{-.5cm}

\newpage

\subsection{Evolution of cooperation in the square lattice}

In~\cite{ohtsuki2006simple}, a general rule for cooperation on graphs was found for the model with one individual per node, where interactions are defined by a specific update rule. The key result of \cite{ohtsuki2006simple} is that under the death-Birth (dB) update rule, natural selection favours cooperators (i.e. mutant individuals here) if $b/c > k$, where $k$ here is the average degree of the graph (i.e., the average number of neighbours per node). No such effect exists under the Bd update rule~\cite{ohtsuki2006simple}. In Appendix \ref{app:pairappx}, we revisit step-by-step the derivation of \cite{ohtsuki2006simple}, which uses a pair approximation, and adapt it to our model. We find that the fixation probability $\Phi_1$ starting from one fully mutant deme, given in Eq.~(\ref{Phi1_pairappx}), does not involve the average degree $k$ and that natural selection never favors cooperation. Again, the key differences are that in our model, birth, death and migration are all independent, and that cooperation occurs within demes.

\begin{figure}[htbp]
    \centering
    \includegraphics[angle=90,origin=c,width=\columnwidth]{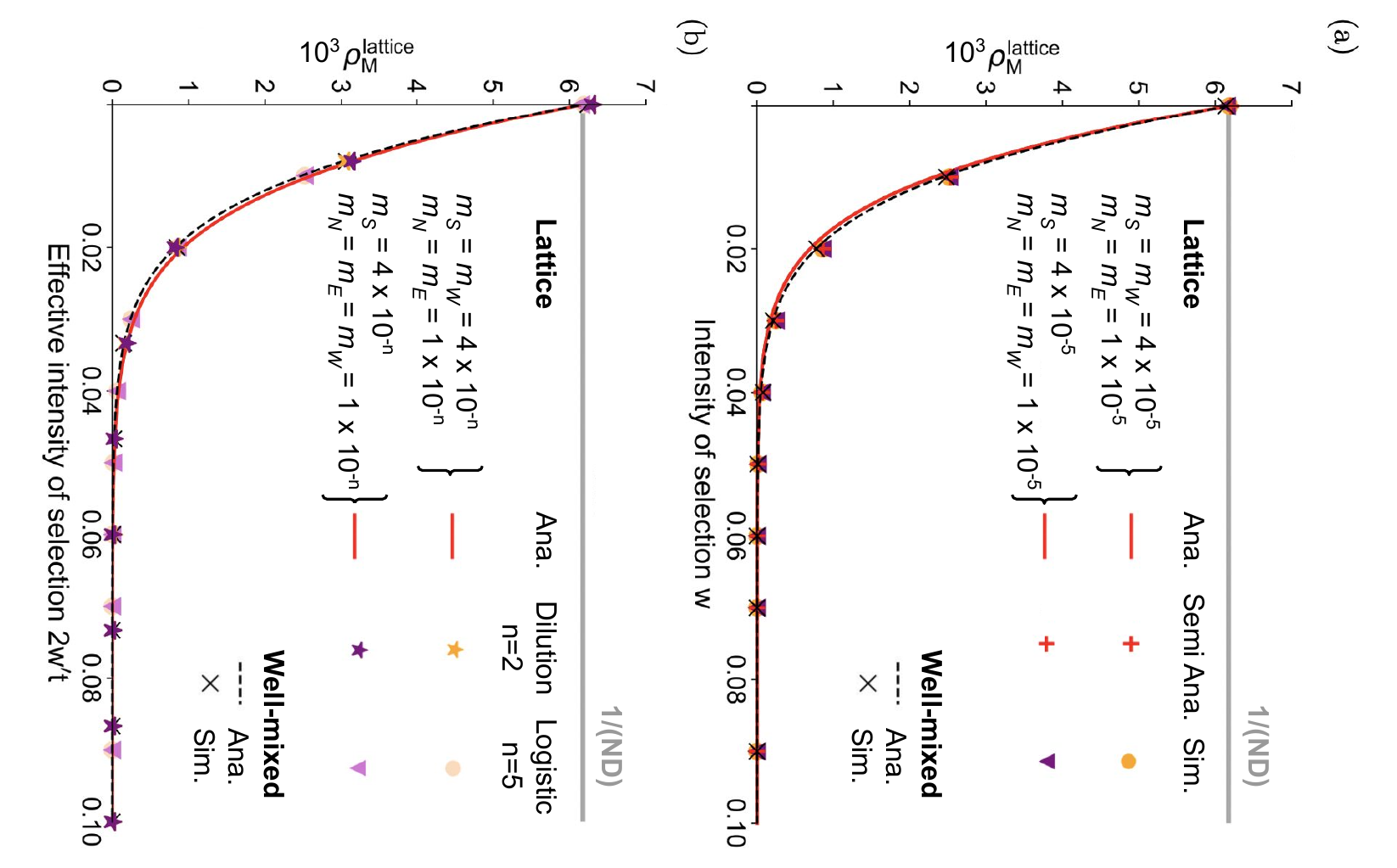} 
    \caption{(a) Mutant fixation probability $\rho^{\textnormal{lattice}}_{M}$ in the square lattice with periodic boundary conditions versus intensity of selection $w$ for different values of migration rates, in the rare migration regime. 
    \\ (b) Mutant fixation probability $\rho^{\textnormal{lattice}}_{M}$ in the square lattice with periodic boundary conditions for the dilution model versus effective intensity of selection $2w't$, for different migration asymmetries and intensities. \\
    Here, we consider $D = 9$ demes with carrying capacity $K = 20$ and steady-state size $N=18$. Other parameter values and conventions are the same as in Fig.~\ref{rhoMcycle}. All simulation results are obtained over $10^6$ replicates. }
    \label{rhoMlattice}
\end{figure}

The proof of~\cite{ohtsuki2006simple} is exact for Bethe lattices, i.e. infinite graphs without cycles where each individual has exactly $k$ neighbors. Graphs possessing this last property are said to be regular. Because Bethe lattices pose some difficulties for simulations, finite regular graphs, including square lattices, were simulated in~\cite{ohtsuki2006simple}, yielding good overall agreement with the prediction that cooperators are favored if $b/c > k$. Thus motivated, here, we study the square lattice with $D=9$ demes shown in Fig.~\ref{structures}, adding periodic boundary conditions, so that each deme is connected to exactly 4 other demes. This graph is a circulation, since for each deme, the incoming migration rates sum to $m_N + m_S + m_E + m_W$, and the outgoing ones too. Therefore, we predict that the fixation probability of a mutant in the rare migration regime is the same as in other circulation graphs (see Section \ref{circs}), and does not depend on migration asymmetry. Fig.~\ref{rhoMlattice}(a) shows that our simulation results are in good agreement with this prediction. In addition, this result seems to extend to frequent migrations in the dilution model, see Fig.~\ref{rhoMlattice}(b).

\newpage
\subsection{Beyond weak selection}
\label{expfit}

So far, we assumed that fitnesses depend linearly on payoffs [see Eq.~(\ref{fit})], and we remained in the weak selection regime where $w\ll 1$. This allowed us to give simple analytical expressions of fixation probabilities in demes, and to conduct a direct comparison with Refs.~\cite{ohtsuki2006simple,ohtsuki2006evolutionary}. 

It is analytically difficult to go beyond weak selection if fitnesses depend linearly on payoffs. However, Ref.~\cite{Traulsen08} showed that calculations beyond the weak selection regime simplify if fitnesses depend exponentially on payoffs. This amounts to writing
\begin{equation}
    \begin{cases}
        f_M(k,N) =e^{\tilde{w}\pi_M(k,N)}\,,\\
        f_W(k,N) =e^{\tilde{w}\pi_W(k,N)}\,,\label{fitexp}
    \end{cases}
\end{equation}
where $\tilde{w}$ is a parameter that determines the intensity of selection (as $w$ in the linear case), instead of Eq.~(\ref{fit}). Payoffs $\pi_M(k,N)$ and $\pi_W(k,N)$ remain expressed by Eq.~(\ref{payoffs}).

Using the same reasoning as before, we obtain the same expressions for the probabilities of fixation of cooperator mutants in subdivided populations for rare migrations, namely Eq.~(\ref{eq_clique_cycle}) for the clique, the cycle, and all circulations, and Eq.~(\ref{phi1starfrom01and10}) for the star. What changes is the expression of the fixation probabilities within demes, which now read~\cite{Traulsen08}
\begin{equation}\label{rhoMexp}
    \rho_M = \frac{1-\exp\left\{\tilde{w}\left[c+b/(N-1)\right]\right\}}{1-\exp\left\{N\tilde{w}\left[c+b/(N-1)\right]\right\}}\,,
\end{equation}
and
\begin{equation}\label{rhoWexp}
    \rho_W = \frac{1-\exp\left\{-\tilde{w}\left[c+b/(N-1)\right]\right\}}{1-\exp\left\{-N\tilde{w}\left[c+b/(N-1)\right]\right\}}\,.
\end{equation}
Their ratio, which enters Eq.~(\ref{eq_clique_cycle})  and Eq.~(\ref{phi1starfrom01and10}), is:
\begin{equation}\label{gammaexp}
    \gamma = \frac{\rho_W}{\rho_M}=\exp\left\{\tilde{w}\left[c(N-1)+b\right]\right\}\,,
\end{equation}
where we neglected the dependence of steady-state deme size on fitness, which is acceptable if $\tilde{w}\ll 1$ (see Methods and Ref.~\cite{marrec2021toward}).
Hence, this exponential convention for fitnesses yields simple analytical expressions of fixation probabilities for rare migrations in our model of spatially structured populations. Here, we assumed $\tilde{w}\ll 1$ but not $N\tilde{w}\ll 1$: in this sense, these results hold beyond weak selection. In addition, the first-order expansions of Eqs.~(\ref{rhoMexp}) and~(\ref{rhoWexp}) when $\tilde{w}\ll 1$ and $N\tilde{w}\ll 1$ exactly match Eqs.~(\ref{rhoMa}) and (\ref{rhoWa}), respectively, with $\tilde{w}$ standing in for $w$. Thus, the exponential and the linear conventions for fitnesses are equivalent in the weak selection regime. 

Importantly, because only the fixation probabilities within demes are modified by changing from linear to exponential fitnesses, our main conclusions remain valid in the exponential case. This includes our results for circulations and stars which are based respectively on Eq.~(\ref{eq_clique_cycle}) and Eq.~(\ref{phi1starfrom01and10}), as well as our reasoning for general graphs based on the pair approximation in Appendix \ref{app:pairappx}. Our main result that natural selection does not favor cooperation in our model of spatially structured populations therefore extends beyond weak selection, at least to the regime where $\tilde{w}\ll 1$ without needing $N\tilde{w}\ll 1$.

\section{Discussion}

In all structures we considered, the fixation probability of a cooperator in a population of defectors is smaller than vice-versa. Thus, cooperation is never favored in our model. The reason for this is that cooperation only takes place within nodes. Migration events do not directly involve cooperation or any interaction between individuals from different demes, not even competition. Cooperation is therefore a local interaction. In contrast, cooperative interactions happen at the level of the graph in \cite{ohtsuki2006evolutionary}. Indeed, each individual (placed on one node) interacts with its nearest neighbors on the graph, which affects its fitness and thus its probability of being picked at the birth step. This separation of scales entails that cooperation cannot be favored by natural selection in our model, including in other graphs, complex networks~\cite{Chen08}, and structures with multiple network layers~\cite{Zhang15}.  

More formally, our model decouples migration, birth, and death events. In the rare migration regime, fixing in one deme and spreading to the whole spatially structured population happen on two different time scales. Because of this, fitnesses enter overall fixation probabilities only through the fixation probabilities in one well-mixed deme, namely $\rho_M$ and $\rho_W$. Assuming that fitnesses depend linearly on payoffs, in the weak selection regime, we can compare Eqs.~(\ref{rhoMa}) and~(\ref{rhoWa}) to the fixation probabilities that would be obtained in the frequency-independent case, i.e. $f_M=f_W(1-\epsilon)$ with a slightly deleterious mutant with fitness disadvantage $\epsilon$ satisfying $0<\epsilon\ll1$, $N\epsilon\ll1$. We get a perfect mapping when choosing $\epsilon=w[c+b/(N-1)]$. Moreover, assuming that fitnesses depend exponentially on payoffs as in Section~\ref{expfit}, Eqs.~(\ref{rhoMexp}) and~(\ref{rhoWexp}) also match the frequency-independent fixation probabilities obtained in that convention with $f_M/f_W=\exp\left\{-\tilde{w}\left[c+b/(N-1)\right]\right\}$, and this holds beyond weak selection, assuming $\tilde{w}\ll 1$ without needing $N\tilde{w}\ll 1$. Thus, the fixation probabilities $\rho_M$ and $\rho_W$ have the same expression as with constant fitnesses, as long as we pick an effective fitness disadvantage for the mutant, which depends on the parameters of cooperation. This confirms that at the level of the whole structure, the frequency-dependence of the model does not matter. Because of this, cooperation cannot benefit from the specificities of the structure. Thus, our previous results regarding the impact of spatial structure on mutant fixation for constant fitnesses~\cite{marrec2021toward,Abbara23} extend to the present model with cooperation.

In our framework, selection is essentially soft, i.e.\ the contributions of demes are not affected by their average fitnesses~\cite{Wallace75}. Indeed, in our dilution model, the total size of each deme after growth does not impact its contribution to the next bottleneck state. In our carrying-capacity model, the carrying capacity of each deme is fixed and independent of its composition, and equilibrium sizes depend only weakly on it. We implemented soft selection because we aimed to assess the impact of population structure in its most minimal form. A promising direction for future research would be to study such a model with hard selection. Indeed, in models where subpopulations contribute to the next bottleneck proportionally to their sizes, the fraction of cooperators can increasing overall despite decreasing within each subpopulation~\cite{chuang2009simpson}: this is known as the Simpson paradox~\cite{blyth1972simpson, simpson1951interpretation, samuels1993simpson, good1987amalgamation}. Such a coupling between composition and population size can favor the evolution of cooperation~\cite{Melbinger10,Cremer12,Cremer19}. It will thus be important to understand how this effect is impacted by graph structure. A first interesting step in this direction was conducted in Ref.~\cite{Fu13}, which determined conditions under which cooperation can be favored at mutation-selection equilibrium, in a model with constant total population size but variable deme sizes.

In this context, we can summarize our results as follows. The ability of specific graphs and update rules to favor the evolution of cooperation in models with one individual per node of the graph~\cite{ohtsuki2006evolutionary,ohtsuki2006simple} does not survive coarse-graining in a model with soft selection. This result does not arise from considering larger deme sizes, but from separating scales. Indeed, in our model, migration is independent from birth and death, and cooperation occurs locally within demes, while spatial structure matters between demes. 

In addition to considering hard selection, possible extensions include addressing other ways through which migration may couple with cooperation~\cite{Wu11}, studying other games such as the snowdrift game~\cite{Gore09}, and explicitly modeling the dynamics of a diffusible resource~\cite{Borenstein13}.
\clearpage
\newpage

\onecolumngrid
\appendix

\section{Formal comparison with models with one individual per node}

\subsection{Cycle \cite{ohtsuki2006evolutionary}}
\label{app:cycle}

In \cite{ohtsuki2006evolutionary}, evolutionary games, and especially the prisoner's dilemma, were studied in the context of the cycle graph with one individual per node and with different update rules (birth-death, death-birth, imitation). Analytical expressions were obtained for the fixation probability of a cooperator in a population of defectors and vice-versa, in the weak-selection regime $w\ll 1$. The key result of \cite{ohtsuki2006evolutionary} is that selection can favour cooperators over defectors for death-birth and imitation updates, but not birth-death updates. To understand the similarities and differences with our model, let us compare the derivations of the fixation probabilities in the cycle within each of the two models. 

In our model with a well-mixed deme on each node of the cycle and migration events independent from birth and death events, and in the rare migration regime, let $i$ denote the number of fully mutant demes. Upon each migration event, the number of mutant demes can either increase by one with probability $\lambda_i$, due to a migration of one mutant individual to a wild-type deme and its subsequent fixation in it, or decrease by one with probability $\mu_i$ upon the opposite succession of events, or stay constant in all other cases. A recurrence relationship can then be written on the fixation probability $\Phi^{\textnormal{cycle}}_{i}$ in a cycle of $D$ demes starting from $i$ consecutive fully mutant demes. For this, one considers all the possibilities at the first migration event that occurs, which allows to relate $\Phi^{\textnormal{cycle}}_{i}$ to $\Phi^{\textnormal{cycle}}_{i-1}$ and $\Phi^{\textnormal{cycle}}_{i+1}$. Solving this recurrence relationship yields
\begin{equation}\label{eq_phi_i_cycle}
    \Phi^{\textnormal{cycle}}_i = \frac{1+\sum_{k=1}^{i-1}\prod_{j=1}^{k}\gamma_j}{1+\sum_{k=1}^{D-1}\prod_{j=1}^{k}\gamma_j}\,,
\end{equation}
where $\gamma_j = \mu_j / \lambda_j=\rho_W / \rho_M$, which is independent of $j$. Therefore, Eq. (\ref{eq_phi_i_cycle}) gives
\begin{equation}\label{eq_phi_i_cycle_2}
    \Phi_i^{\textnormal{cycle}} = \frac{1-\gamma^i}{1-\gamma^D}, \quad \textnormal{with} \quad \gamma = \frac{\rho_W}{\rho_M}\,.
\end{equation}
A more detailed derivation (which is presented without frequency-dependent fitness, but extends to the present case) is given in~\cite{marrec2021toward}, and the general method is presented in~\cite{traulsen2009stochastic}. Note that the expression in Eq. (\ref{eq_phi_i_cycle_2}) also holds in the case of the clique, where it can be obtained with an extremely similar derivation~\cite{marrec2021toward}.

In the model with one individual per node of a cycle with $N$ nodes considered in \cite{ohtsuki2006evolutionary}, the same approach was employed to express the fixation probability $\rho_M^{\textnormal{cycle}}$ of one mutant, yielding
\begin{equation}\label{eq_phi_i_cycle_ON}
    \rho_M^{\textnormal{cycle}} = \frac{1}{1+\sum_{k=1}^{N-1}\prod_{j=1}^{k}\gamma_j}\,,
\end{equation}
which is formally equivalent to $\Phi^{\textnormal{cycle}}_{1}$ in our model (see Eq. (\ref{eq_phi_i_cycle})). However, in this model, $\gamma_j = \mu_j / \lambda_j$, where $\lambda_j$ (resp.  $\mu_j$) represents the probability that the number of mutant nodes increases (resp. decreases) by one upon an update, takes different expressions depending on the update rule~\cite{ohtsuki2006evolutionary}. Indeed, with the birth-death update rule, 
\begin{equation} \label{BDcycle}
\gamma_j=\frac{f_W^e}{f_M^e}=\frac{1-w+wb}{1-w+w(b-2c)}\,,
\end{equation}
for $2\leq j\leq N-2$, where $f_W^e$ (resp. $f_M^e$) is the fitness of a wild-type (resp. mutant) at the edge of the cluster of mutant nodes, i.e. which has a mutant neighbor and a wild-type neighbor. Here we expressed each fitness as $1-w+w\pi$ where the total payoff $\pi$ received by an individual on the cycle is computed by summing the payoffs they receive from interactions with their two neighbors, with the payoff matrix in Eq.~(\ref{payoffmatrix}). We also used the fact that during the fixation process, starting from one mutant, there is a cluster of consecutive mutant nodes that cannot break. With the death-birth update rule, the fitnesses involved in $\gamma_j$ depend on whether the mutant that was at an edge of the cluster of mutant nodes (individual 1) or its wild-type neighbor (individual 2) is selected to die. If individual 1 is selected to die, which can lead to a decrease of $j$ by one, the fitness of the wild-type (resp. mutant) neighbor of individual 1 is denoted by $f_W^e$ (resp. $f_M^e$). If individual 2 is selected to die, which can lead to an increase of $j$ by one, the fitness of the wild-type (resp. mutant) neighbor of individual 1 is denoted by $f_W^{'e}$ (resp. $f_M^{'e}$). We have
\begin{equation} \label{DBcycle}
\gamma_j=\frac{f_W^e}{f_W^e+f_M^e}\frac{f_W^{'e}+f_M^{'e}}{f_M^{'e}}=\frac{1-w+wb}{2-2w+w(3b-2c)}\frac{2-2w+w(b-2c)}{1-w+w(b-2c)}\,,
\end{equation} 
for $3\leq j\leq N-3$. The expressions in Eqs. (\ref{BDcycle}) and (\ref{DBcycle}) differ for all $b>c>0$, which leads to different probabilities of fixation under these two different update rules~\cite{ohtsuki2006evolutionary}.

This comparison of the derivation of fixation probabilities within our model and within that of~\cite{ohtsuki2006evolutionary} shows that the key difference is that in our model, migration events do not depend on birth or death and are thus decoupled from interactions between individuals, while in models with one individual per node, any move on the graph is coupled to birth and death via the update rule. 

\subsection{General graphs with pair approximation \cite{ohtsuki2006simple}}
\label{app:pairappx}

While~\cite{ohtsuki2006evolutionary} focused on the cycle, \cite{ohtsuki2006simple} derived a general rule for cooperation on graphs in the same model with one individual per node. The key result of \cite{ohtsuki2006simple} is that under the death-birth (dB) update rule, natural selection favours cooperators (i.e. mutant individuals here) if $b/c > k$, where $k$ here is the average degree of the graph (i.e., the average number of neighbours per node). No such effect exists under the Bd update rule~\cite{ohtsuki2006simple}. To understand the origin of the difference between these results and ours, we here follow the main analytical derivation of \cite{ohtsuki2006simple} and adapt it to our model. Again, the key difference is that our model does not involve any update rule, as birth, death and migration are all independent. We will point out where this matters.  

Following~\cite{ohtsuki2006simple}, let $p_M$ (resp. $p_W$) denote the frequency of fully mutant $M$ (resp. wild-type $W$) demes in a given graph comprising $D$ nodes. In our model, each of these nodes contains a well-mixed deme, while it comprises a single individual in~\cite{ohtsuki2006simple}. Let us also introduce the frequencies $p_{MM}$, $p_{WW}$, $p_{MW}$ and $p_{WM}$ of neighbouring $MM$, $WW$, $MW$ and $WM$ pairs of demes, and the conditional probability $q_{X|Y}$ to find an $X$ deme at a certain position given that one randomly chosen nearest-neighbour of this $X$ deme (i.e. a deme connected to it by an edge) is a $Y$ deme. We focus on the rare migration regime. As in \cite{ohtsuki2006simple}, we will make a \textit{pair approximation} where only the pair correlation is accounted for, meaning that higher-order correlations are fully determined by these pair correlations. The pair approximation requires Bethe lattices (or Cayley trees), which are regular graphs (i.e., all nodes have the same degree) with no loops. We thus focus on regular graphs, and we further assume $k>2$ (the case $k=2$ corresponds to the cycle, treated in the previous section).
The following identities hold:
\begin{equation}
\begin{cases}
    p_M + p_W = 1\,, \\
    q_{M|X}+q_{W|X} = 1\,, \\
    p_{XY} = q_{X|Y}\,p_Y\,, \\
    p_{MW} = p_{WM}\,.
\end{cases}
\end{equation} 
Thus, the system is fully described by $p_M$ and $q_{M|M}$. We will focus on these two quantities.

To derive the fixation probability of a mutant starting from a fully mutant deme (and $D-1$ fully wild-type demes), we first compute the probabilities that $p_M$ changes upon one migration event~\cite{ohtsuki2006simple}. The probability that the total number of $M$ demes increases by one (meaning that one $M$ individual has migrated from a fully $M$ deme to a fully $W$ deme and has fixed there) is given by:
\begin{equation}
    P\left(\Delta p_M = \frac{1}{D}\right) = p_W \sum_{k_M=0}^k \binom{k}{k_M} q^{k_M}_{M|W}\, q^{k-k_M}_{W|W}\, \frac{k_M}{k}\rho_M\,,
\end{equation}
where $k$ is the degree of our regular graph, while $k_M$ (resp. $k-k_M$) is the number of $M$ (resp. $W$) neighbours of a deme. Indeed, for this to happen, a $W$ deme needs to be picked as the deme of arrival of the migration (probability $p_W$), one of its $M$ neighbors needs to be picked as the deme of origin of the migration (probability $k_M/k$), and the mutant migrant has to fix in the wild-type deme (probability $\rho_M$). In addition, we need to take into account all different possibilities for the numbers of neighbors of each type that the $W$ deme where the migration arrives possesses, leading to a sum with binomial weights. Recognising a binomial sum, we obtain
\begin{equation}
    P\left(\Delta p_M = \frac{1}{D}\right) = p_W q_{M|W}\rho_M=p_{MW}\rho_M\,, 
\end{equation}
which has a simple interpretation: we need to pick a $WM$ neighbouring pair for our migration event, and then the mutant needs to fix in the wild-type neighboring deme. Importantly, this last simplification cannot be made in the proof of \cite{ohtsuki2006simple}, because the replacement probabilities upon an update (which are replaced by fixation probabilities $\rho_M$ in our model) depend upon the state of the neighborhood (i.e., on the value of $k_M$) due to the cooperative interaction. In the model of \cite{ohtsuki2006simple}, this dependence involves the update rule, and vanishes for the Bd update rule while it remains for the dB one. In fact, here, the reasoning we just made starts by picking the arrival deme of the migration (in a dB spirit), but an equivalent one can be made by first picking the deme where the migration originates (in a Bd spirit):
\begin{equation}\label{plus1}
    P\left(\Delta p_M = \frac{1}{D}\right) = p_M \sum_{k_M=0}^k \binom{k}{k_M}q^{k_M}_{M|M}\, q^{k-k_M}_{W|M}\, \frac{k-k_M}{k}\rho_M= p_{MW}\rho_M\,. 
\end{equation}
Indeed, a $M$ deme needs to be picked as the deme of origin of the migration (probability $p_M$), one of its $W$ neighbors needs to be picked as the deme of arrival of the migration (probability $(k-k_M)/k$), and the mutant migrant has to fix in the wild-type deme (probability $\rho_M$), and we need to take into account all different possibilities for the numbers of neighbors of each type that the $M$ deme where the migration originates possesses.

Similarly, the probability that the total number of $M$ demes decreases by one (meaning one $W$ individual has migrated from a fully $W$ deme to a fully $M$ deme and has fixed there) reads:
\begin{equation}\label{minus1}
    P\left(\Delta p_M = -\frac{1}{D}\right) =p_{MW}\rho_W\,.
\end{equation}
Upon one migration event, only the two types of events discussed above can change the state of the population. Either the total number of $M$ demes increases by one with probability in Eq.~(\ref{plus1}) or it decreases it by one with probability in Eq.~(\ref{minus1}). Therefore, the expectation of the variation $\Delta p_M$ of $p_M$ upon a migration event reads:
\begin{equation}
\begin{split}
    \left\langle \Delta p_M \right\rangle & = \frac{1}{D}P\left(\Delta p_M = \frac{1}{D}\right) + \left(-\frac{1}{D}\right)P\left(\Delta p_M = -\frac{1}{D}\right) \\
    & = \frac{1}{D} p_{MW} \left(\rho_M-\rho_W\right)=\frac{1}{D} \left(1-q_{M|M}\right)p_M \left(\rho_M-\rho_W\right) \,.
\end{split}
\end{equation}
In the weak selection regime $w\ll 1$, we can compute a Taylor expansion of the two fixation probabilities $\rho_M$ and $\rho_W$ (see Eqs. (\ref{rhoM}) and (\ref{rhoW})). $\left\langle \Delta p_M \right\rangle$ can then be simplified as:
\begin{equation}
    \left\langle\Delta p_M \right\rangle =  -\frac{w}{ND}\left[b+c(N-1)\right]\left(1-q_{M|M}\right)p_M + O(w^2)\,.
\end{equation}
In the deterministic limit of a very large number of demes, the evolution of $p_M$ is therefore described by:
\begin{equation} 
    \frac{dp_M}{dt} =  -\frac{w}{ND}\left[b+c(N-1)\right]\left(1-q_{M|M}\right)p_M + O(w^2) = w\, F_1\left(p_M, q_{M|M}\right) + O(w^2)\,, 
\end{equation}
where we introduced
\begin{equation} 
    F_1\left(p_M, q_{M|M}\right) =  -\frac{1}{ND}\left[b+c(N-1)\right]\left(1-q_{M|M}\right)p_M\,, 
\end{equation}
and where the unit of time corresponds to one migration step.

To fully describe the dynamics of the system in the deterministic regime and within the pair approximation, we also need to write an equation on the conditional probability $q_{M|M}$~\cite{ohtsuki2006simple}. For this, let us focus on the joint probability $p_{MM}$, and write down the probabilities that the number of $MM$ pairs in the graph increases or decreases upon a migration event. Let us reason in a dB spirit, i.e. by first choosing the arrival deme of the migration (here again, reasoning in a Bd spirit gives the exact same result). For $p_{MM}$ to increase, the arrival deme needs to be a $W$ deme, which occurs with probability $p_W$, and a migration needs to arrive there from a neighboring $M$ deme, the associated probability being $k_M/k$, where $k_M$ is the number of $M$ neighbors of the arrival $W$ deme. Then, if the mutant fixes, which occurs with probability $\rho_M$, $p_{MM}$ increases by $2k_M/(kD)$. Thus,
\begin{equation}
    P\left(\Delta p_{MM}=\frac{2k_M}{kD}\right)=p_W\binom{k}{k_M}q_{M|W}^{k_M}q_{W|W}^{k-k_M}\frac{k_M}{k}\rho_M\,,
\end{equation}
and similarly
\begin{equation}
    P\left(\Delta p_{MM}=-\frac{2k_M}{kD}\right)=p_M\binom{k}{k_M}q_{M|M}^{k_M}q_{W|M}^{k-k_M}\frac{k-k_M}{k}\rho_W\,.
\end{equation}
The expectation of $\Delta p_{MM}$ thus reads:
\begin{equation}
\begin{split}
    \left\langle\Delta p_{MM} \right\rangle & =  \sum^k_{k_M = 0} \frac{2k_M}{kD}P\left(\Delta p_{MM} = \frac{2k_M}{kD}\right) - \sum^k_{k_M = 0} \frac{2k_M}{kD}P\left(\Delta p_{MM} = - \frac{2k_M}{kD}\right) \\
    & = \frac{2}{kND}p_M\left(1-q_{M|M}\right)\left[1+\left(k-1\right) \frac{p_M-q_{M|M}}{1-p_M}\right] + O(w)\,.
\end{split}
\end{equation}

In the deterministic limit, we have:
\begin{equation}
\begin{split}
\frac{dq_{M|M}}{dt} & = \frac{d}{dt}\left( \frac{p_{MM}}{p_M}\right) =  \frac{1}{p_M}\frac{dp_{MM}}{dt} + O(w) \\
& =  \frac{2}{kND}\left(1-q_{M|M}\right)\left[1+\left(k-1\right) \frac{p_M-q_{M|M}}{1-p_M}\right] + O(w) \\
& = F_2\left(p_M, q_{M|M}\right) + O(w)\,,
\end{split}
\end{equation}
where we introduced
\begin{equation}
    F_2\left(p_M, q_{M|M}\right)= \frac{2}{kND}\left(1-q_{M|M}\right)\left[1+\left(k-1\right) \frac{p_M-q_{M|M}}{1-p_M}\right]\,.
\end{equation}

By computing the probabilities that $p_M$ and $p_{MM}$ change in one migration step, we derived two differential equations on our two variables of interest, $p_M$ and $q_{M|M}$, in the deterministic limit. In the weak selection regime, we suppose that the conditional probability $q_{M|M}$ which describes the local interactions reaches an equilibrium much more quickly than the frequency $p_M$ (which describes the overall state of the structure)~\cite{ohtsuki2006simple}. Therefore, the system quickly converges onto the space defined by $F_2\left(p_M, q_{M|M}\right) = 0$. This gives
\begin{equation}
    q_{M|M} = \frac{k-2}{k-1}p_M + \frac{1}{k-1}
    \,. \label{rln}
\end{equation}
Under this assumption, the system is fully described by $p_M$. 

In order to compute probabilities of fixation, we need to go beyond the deterministic limit, and to work in the diffusion limit~\cite{ohtsuki2006simple}. To write a diffusion equation satisfied by $p_M$, we need the second moment $\left\langle \Delta p_M^2 \right\rangle$ of $\Delta p_M$, which reads
\begin{equation}
\begin{split}
    \left\langle\Delta p_M^2 \right\rangle & = \frac{1}{D^2}P\left(\Delta p_M = \frac{1}{D}\right) + \frac{1}{D^2}P\left(\Delta p_M = - \frac{1}{D}\right) \\
    & = \frac{1}{D^2}p_{MW}(\rho_M+\rho_W) = \frac{2}{ND^2}p_{MW} + O(w) \\
    & = \frac{2}{ND^2}p_M\left(1-q_{M|M}\right) + O(w) \\
    & = \frac{2}{ND^2}\frac{k-2}{k-1}\left(1-p_M\right)p_M + O(w)\,,
\end{split}
\end{equation}
where we have used Eq.~(\ref{rln}). We can now write down a Kolgomorov backward equation on the fixation probability $\phi_{M}(y)$ of $M$ starting with an initial frequency $p_M(t = 0) = y$: 
\begin{equation}\label{kolgomoroveq} 
    m(y)\frac{d\phi_M(y)}{dy} + \frac{v(y)}{2}\frac{d^2\phi_M(y)}{dy^2} = 0\,,
\end{equation}
where $m(y)$ and $v(y)$ are given by:
\begin{equation}
    m(y) = -\frac{w}{ND}\frac{k-2}{k-1}\left[b+c(N-1)\right]\left(1-y\right)y \quad \textnormal{and} \quad
    v(y) = \frac{2}{ND^2}\frac{k-2}{k-1}\left(1-y\right)y\,.
\end{equation}
Recall that $k > 2$ and that we focus on the weak selection regime $w \ll 1$. This Kolgomorov backward equation can be solved with the two conditions $\phi_{M}(0) = 0$ and $\phi_{M}(1) = 1$ (which describe the two absorbing states):
\begin{equation}
    \phi_M(y) = y - \frac{wD}{2}\left[b+c(N-1)\right]y\left(1-y\right) + O(w^2)\,.
\end{equation}
Starting from one single fully mutant deme, the probability that mutants take over the population is thus:
\begin{equation}\label{Phi1_pairappx}
        \Phi_1\equiv\phi_M\left(\frac{1}{D}\right) = \frac{1}{D}\left\{ 1-\frac{w}{2}\left[b+c(N-1)\right]\left(D-1\right)\right\} + O(w^2) \,.
\end{equation}
Similarly, starting from one single fully wild-type deme, the probability that wild-types take over the population is:
\begin{equation}
        \Phi_{1,W}\equiv1-\phi_M\left(\frac{D-1}{D}\right) = \frac{1}{D}\left\{ 1+\frac{w}{2}\left[b+c(N-1)\right]\left(D-1\right)\right\} + O(w^2) \,.
\end{equation}
Thus, in the weak-selection limit, for all $b > c > 0$, we have
\begin{equation}
        \Phi_1 < \frac{1}{D} <\Phi_{1,W}\,,
\end{equation}
where $1/D$ is the neutral fixation probability. Therefore, natural selection never favors cooperation in our model. This stands in constrast to the result of \cite{ohtsuki2006simple}, where, under the dB update rule, the fixation probability of a single (cooperator) mutant exceeds that of a single (defector) wild-type organism if $b/c<k$. As mentioned above in the case of the cycle, this is because cooperation does not come in at the same level in both models. In \cite{ohtsuki2006simple}, each individual (placed on one node) interacts with its nearest neighbors on the graph. Cooperative interactions thus happen at the level of the structure. Conversely, our model, cooperative effects only occur within a deme.

\section{Details on the serial dilution model}
\label{dilution_details}

\subsection{Model for structured populations on graphs}
The dilution model is directly inspired from~\cite{Abbara23}, but adds cooperation through the payoff matrix in Eq.~(\ref{payoffmatrix}). We consider $D$ demes located on the nodes of a graph. Migrations can happen along the edges of the graph, according to a matrix $(m_{ij})_{i,j=1,\dots,D}$, where $m_{ij}$ is the migration probability from deme $i$ to deme $j$ at each bottleneck. To ensure conservation of the total number of individuals at each bottleneck, migration probabilities satisfy $\sum_j m_{ji} = 1$ for all $i=1,\dots,D$. The population undergoes successive bottlenecks where all demes have size $B$, (that we take equal to $N$, the steady-state size in the carrying-capacity model), following a 2-step process:
\begin{enumerate}
    \item A local growth phase with cooperation occurs in each deme, as explained in Section~\ref{dilution_model}. We denote by $k'_i(0)$ (resp.\ $l'_i(0)$) the number of mutant (resp.\ wild-type) individuals that are in deme $i$ at a given bottleneck, satisfying $k'_i(0)+l'_i(0)=N$. These numbers grow following Eq.~(\ref{growth_phase}) during a fixed time $t$. At the end of the growth phase, deme $i$ contains $k'_i(t)$ mutants and $l'_i(t)$ wild-types. Note that because the growth phase is modeled via ordinary differential equations, these numbers are not necessarily integers. The fraction of mutants at the end of the growth phase in deme $i$ is $x_i= k'_i(t)/[k'_i(t)+l'_i(t)]$.
    \item A dilution and migration step then occurs, where we sample the incoming individuals to a deme $i$ from a multinomial distribution with $N$ trials and with probabilities $x_j m_{ji}$ (resp.\ $(1-x_j) m_{ji}$) to sample a mutant (resp.\ wild-type) from deme $j$, for all $j\in \{1,\dots,D\}$. Sampling from this law, we get numbers $k'_{ji}$ (resp.\ $l'_{ji}$) of mutants (resp.\ wild-types) sent to deme $i$ from each deme $j$. Thus, the total number of mutants in deme $i$ at the new bottleneck is $k'_i=\sum_{j=1}^D k'_{ji}$, and the total number of wild-types is $l'_i=\sum_{j=1}^D l'_{ji}$. These two numbers sum to $N$, due to the properties of multinomial sampling. They play the part of the numbers $k'_i(0)$ and $l'_i(0)$ introduced above for the next iteration of steps 1 and 2. 
\end{enumerate}

\subsection{Fixation probability in a deme in the dilution model}
\label{dilution_wellmixed}
Let us consider a single well-mixed deme, which corresponds to the previous case with $D=1$. The multinomial distribution then turns into a binomial law, and the serial dilution model thus becomes similar to a Wright-Fisher model. To calculate the fixation probability of a mutant, we can use Kimura's diffusion approximation~\cite{CrowKimura}, in the regime of large population size $N$ and small intensity of selection $w't$, as detailed in~\cite{Abbara23} for mutants with constant fitness. To adapt this approach to cooperator mutants, we need an analytical expression of the fraction of mutants $x(t)=k'(t)/[k'(t)+l'(t)]$ at the end of a growth phase, starting from a fraction $x(0)\equiv x_0$ at the bottleneck. During the growth phase, i.e. for $\tau\in[0,t]$, the numbers $k'$ and $l'$ grow according to Eq.~(\ref{growth_phase}), which yields:
\begin{align}
\label{growthDil}
    \dfrac{d k'(\tau)}{d\tau}&= k'(\tau)  \left[ 1 - w' + w' \left( b\,\dfrac{k'(\tau)-1}{k'(\tau) + l'(\tau) - 1}-c \right)\right] \,, \\
    \dfrac{d l'(\tau)}{d\tau}&= l'(\tau)  \left[ 1 - w' + w'b\, \dfrac{ k'(\tau)}{k'(\tau) + l'(\tau) - 1} \right]\,.
\end{align}
With parameters $b,c$ of order 1 and $N \gg 1$, we can neglect the $-1$ in the denominators in the two equations above, which then yield:
\begin{equation}
    \dfrac{k'(t)}{l'(t)} = \dfrac{k'(0)}{l'(0)} \exp(-w'ct).
    \label{evolratiokl}
\end{equation}
Note that the same equation holds with mutants with constant fitness~\cite{Abbara23}, where wild-types have fitness $f_W=1$, and mutants have fitness $f_M=1+s$, but with $s$ instead of $-w'c$. Thus, here, with the approximations we made, $-w'c$ plays the part of a constant fitness advantage. The evolution of the mutant fraction $x=k'/(k'+l')$ can be directly deduced from Eq.~(\ref{evolratiokl}). Under Kimura's diffusion approximation, the fixation probability $\rho'(x_0)$ starting from a fraction $x_0$ of mutants in a population of size $N\gg1$, with $w'\ll 1$, then reads:
\begin{equation}
    \rho'(x_0)=\dfrac{1-e^{2Nw'ctx_0}}{1-e^{2Nw'ct}}.
    \label{rho_kimura}
\end{equation}

This yields the probability $\rho_M$ of fixation of a single mutant 
\begin{equation}
\label{rhoM_dil}
    \rho_M'\equiv \rho'(1/N) = \dfrac{1 - e^{2 w'ct}}{1-e^{2 N w'ct}}.
\end{equation}
The probability $\rho'_W$ of fixation of a single wild-type individual in a population where other individuals are mutants can be obtained similarly and reads:
\begin{equation}
\label{rhoW_dil}
    \rho_W'= \dfrac{1 - e^{-2 w'ct}}{1-e^{-2 N w'ct}}.
\end{equation}
We can use these expressions to calculate the fixation probability in a structure in the rare migration regime, via Eq.~(\ref{pfix_gal}). 

\subsection{Mapping between carrying-capacity and dilution model in the diffusion approximation}
\label{map_models}

For simplicity, let us first consider the case where individuals have constant fitness. In our work, evolution within a deme is modeled via the Moran model in the carrying-capacity model (neglecting deme size fluctuations), and via a dilution model close to the Wright-Fisher model, where the $e^{s't}-1\approx s't$ plays the role of the fitness advantage $s_\mathrm{WF}$ in the Wright-Fisher model~\cite{Abbara23}. In a well-mixed population of constant size $N$, the mutant fixation probability starting from a fraction $x_0$ of mutants is well-known in the diffusion limit, both for the Moran and the Wright-Fisher models~\cite{kimura64,Blythe2007,CrowKimura}. Let $f_W=1$ denote the wild-type fitness, $s$ the mutant's fitness advantage in the Moran model and $s'$ in the Wright-Fisher model. If $N\gg1$ and $s, s' \ll 1$, the diffusion approximation predicts a fixation probability
\begin{equation}
    \rho^{\textnormal{diff.}}(x_0) = \dfrac{1- e^{N\sigma x_0}}{1-e^{N\sigma}} \,,
\end{equation}
where $\sigma=s$ in the Moran model, while $\sigma=2s_\mathrm{WF}$ in the Wright-Fisher model, and $\sigma=2(e^{s't}-1)\approx 2s't$ in our dilution model. Thus, there is a mapping between $s$ in the carrying-capacity model and $2s't$ in the dilution model. Note that the factor of 2 differs between the Moran model and the Wright-Fisher model: it arises from their different variances in offspring number~\cite{Ewens79}.

Let us now include cooperation. In our serial dilution model, the probability of fixation of a single cooperator mutant is given by Eq.~(\ref{rhoM_dil}) in the diffusion limit.
Let us write the first-order expansion of the previous equation in $w'$, assuming that $w'ct \ll 1$ and $w'ct N\ll 1$:
\begin{equation}
    \rho_M'\approx \dfrac{1}{N}[1 - (N-1) w'ct] + o(w').
\end{equation}
In the regime of parameters we consider here, where $N\gg1$, while $b$ and $c$ are of order 1, this is the same as the probability of fixation $\rho_M$ obtained in Eq.~(\ref{rhoMa}) within the carrying-capacity model, but replacing $w$ by $2w't$. Thus, $2w't$ is the effective selection intensity we use in the dilution model for our quantitative comparisons with the carrying-capacity model.

\section{Variant of the carrying capacity model}
\label{OtherModel}

In the logistic model used throughout, the division rate of individuals of type $A=M,W$ in a deme is given by the logistic function $f_A(1 - N/K)$, where $f_A$ is the fitness given in Eq.~(\ref{fit}) and $N$ is the number of individuals in the deme. Meanwhile, the death rate $g$ is independent from type and deme size (see Model and methods). An alternative way of implementing logistic regulation of population size is to consider division rates that do not depend on deme size but a death rate that does. Let us consider such a variant of the carrying capacity model, by taking a division rate $f_A$ for individuals of type $A=M,W$ and a death rate $\bar{f}N/K$, where $\bar{f}$ is the mean fitness in the deme, for all individuals. Since $\bar{f}=\left[kf_M(k,N)+(N-k)f_W(k,N)\right]/N$, the death rate reads $\left[kf_M(k,N)+(N-k)f_W(k,N)\right]/K$. In this model, the steady state of the deme is equal to its carrying capacity $K$, which thus plays the role of $K(1-g/f)$ in our usual model.

\begin{figure}[htbp]
    \centering
    \includegraphics[width=0.45\columnwidth]{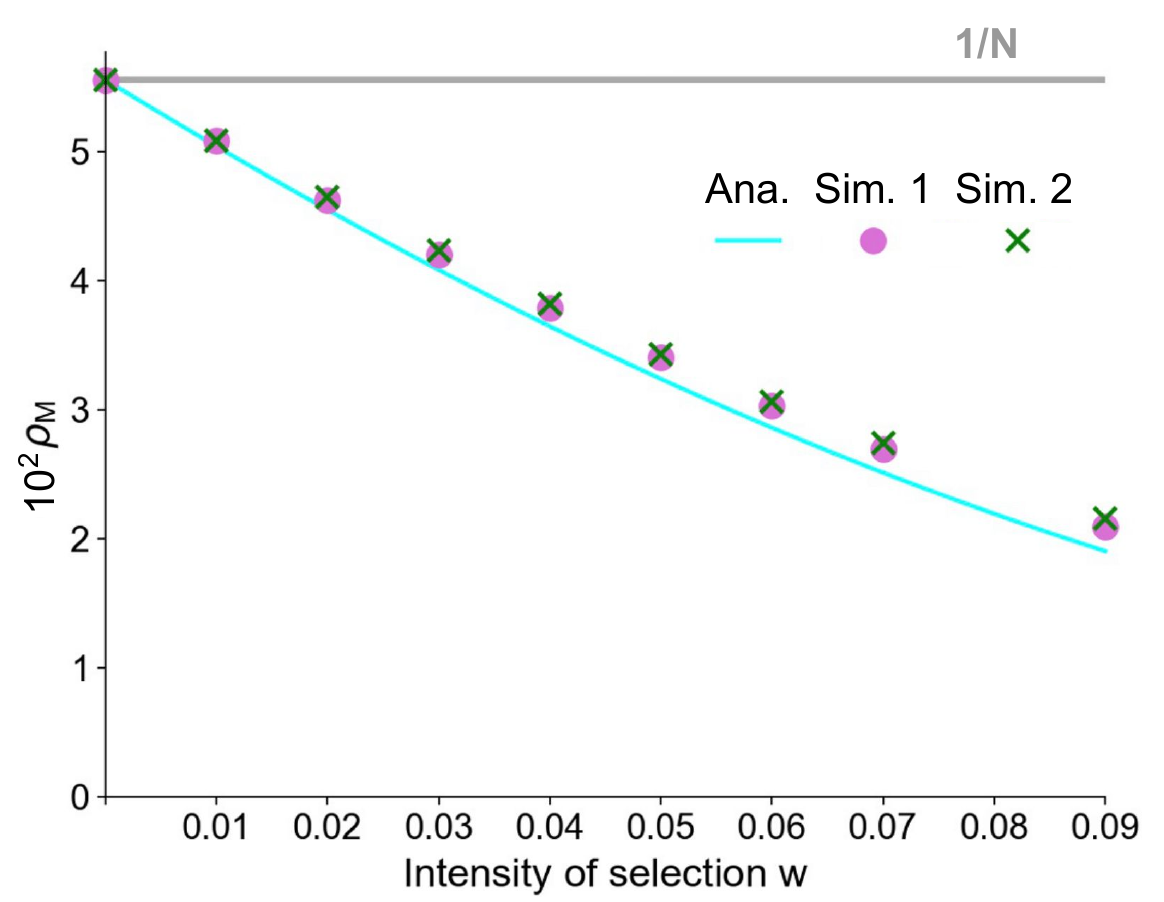} 
    \caption{Mutant fixation probability $\rho_{M}$ in a deme versus intensity of selection $w$. The cyan solid line represents analytical predictions; markers are simulation results obtained for two variants of the carrying capacity model. The first one (``Sim. 1'') is the one used throughout this work, while the second one (``Sim. 2'') is the one presented here, where the logistic regulation is implemented in the death rate and not in the birth rate. The neutral mutant case (gray line) is shown for reference. Analytical predictions are obtained using Eq.~(\ref{rhoM}) that holds for fixed deme size. Simulation results are obtained over $10^7$ replicates.  
    In the first model, we consider a deme with carrying capacity $K = 20$, death rate $g=0.1$ yielding steady-state size $N=18$. In the second model, we consider a deme with carrying capacity $K=18$. In both cases, benefit and cost are $b = 2$ and $c = 1$.}
    \label{TwoModels}
\end{figure}

In the rare migration regime, within-deme events enter the fixation probabilities in graph-structured populations only through the fixation probabilities in demes. Therefore, to compare the two variants of the carrying capacity models in this regime, it is sufficient to compare the fixation probabilities in demes. In Fig.~\ref{TwoModels}, we present simulation results corresponding to these two variants, which are in good agreement. As expected, they are also both close to the analytical prediction obtained within the Moran model. The slight discrepancy arises from deme size fluctuations, which do not exist in the Moran model. Note however that the two models yield different time scales. For instance, in the simple case where $f_W=f_M=1$, at steady state a division event occurs every $1/g$ time unit on average in the first model, and every time unit in the second model.

\section{Numerical simulation methods}
\label{App_simu}

In the carrying-capacity model, the Gillespie algorithm \cite{gillespie1976general,gillespie1977exact} allows us to exactly simulate the stochastic evolution of our population, without requiring any time discretization. In a population with $D$ demes indexed by $i$, and with migration rates $m_{ij}$ from deme $i$ to deme $j$, where two types, $W$ and $M$, exist, the possible elementary events (or ``reactions'') are, for all $i$ and $j$:
\begin{itemize}
    \item $M_i \xrightarrow{k^{+}_M} M_i + M_i$ : birth of one mutant in deme $i$ with rate $k^{+}_M = f_M\left[ 1 - (N_{W,i}+N_{M,i})/K\right]$;
    \item $W_i \xrightarrow{k^{+}_W} W_i + W_i$ : birth of one wild-type in deme $i$ with rate $k^{+}_W = f_W\left[ 1 - (N_{W,i}+N_{M,i})/K\right]$;
    \item $M_i \xrightarrow{k^{-}_M} \emptyset$ : death of a mutant in deme $i$ with rate $k^{-}_M = g_M$;
    \item $W_i \xrightarrow{k^{-}_W} \emptyset$ : death of a wild-type in deme $i$ with rate $k^{-}_W = g_W$;
    \item $M_i \xrightarrow{m_{ij}} M_j$ : migration of a mutant from deme $i$ to deme $j$;
    \item $W_i \xrightarrow{m_{ij}} W_j$ : migration of a wild-type from deme $i$ to deme $j$.
\end{itemize} 
Here, we have denoted by $N_{M,i}$ and $N_{W,i}$ the numbers of mutant and wild-type individuals in deme $i$. The total rate of possible events is then given by:
\begin{equation}\label{Ktot}
    k_{tot} = \sum_{i=1}^D\left[(k^+_{W,i}+k^-_{W,i})N_{W,i} + (k^+_{M,i}+k^-_{M,i})N_{M,i}\right] + \sum_{i,j=1}^{D}m_{ij}(N_{W,i}+N_{M,i})\,.
\end{equation}
Reactions occur at time intervals picked in an exponential distribution with mean $1/k_{tot}$, and the specific reaction that occurs is selected randomly proportionally to the ratio of its rate and $k_{tot}$. Reactions are performed until one of the two types takes over the entire population.

In the dilution model, we performed our simulations as described in~\cite{Abbara23}, except that multinomial sampling was employed throughout, as described in Appendix \ref{dilution_details}, and that the deterministic growth was performed using a numerical resolution of Eq.~(\ref{growthDil}) to include cooperation.

\section*{Code availability}
Our code is freely available in the GitHub repository \url{https://github.com/Bitbol-Lab/Cooperation_structured_pop}.

\section*{Acknowledgments}
This project has received funding from the European Research Council (ERC) under the European Union’s Horizon 2020 research and innovation programme (grant agreement No.~851173, to A.-F.~B.).

\bibliographystyle{unsrt} 
\bibliography{twocylinders} 

\begin{thebibliography}{10}

\bibitem{Hunter91}
T.~Hunter.
\newblock Cooperation between oncogenes.
\newblock {\em Cell}, 64(2):249--270, 1991.

\bibitem{axelrod81}
R.~Axelrod and W.~D. Hamilton.
\newblock The evolution of cooperation.
\newblock {\em Science}, 211(4489):1390--1396, 1981.

\bibitem{celiker2013cellular}
H.~Celiker and J.~Gore.
\newblock Cellular cooperation: insights from microbes.
\newblock {\em Trends in cell biology}, 23(1):9--15, 2013.

\bibitem{mitchison2004t}
N.~A. Mitchison.
\newblock T-cell--{B}-cell cooperation.
\newblock {\em Nature Reviews Immunology}, 4(4):308--312, 2004.

\bibitem{dugatkin1997cooperation}
L.~A. Dugatkin.
\newblock {\em Cooperation among animals: an evolutionary perspective}.
\newblock Oxford University Press, USA, 1997.

\bibitem{raihani2012punishment}
N.~J. Raihani, A.~Thornton, and R.~Bshary.
\newblock Punishment and cooperation in nature.
\newblock {\em Trends in ecology \& evolution}, 27(5):288--295, 2012.

\bibitem{chase1980cooperative}
I.~D. Chase.
\newblock Cooperative and noncooperative behavior in animals.
\newblock {\em The American Naturalist}, 115(6):827--857, 1980.

\bibitem{dugatkin2000cheating}
L.~A. Dugatkin.
\newblock {\em Cheating monkeys and citizen bees: the nature of cooperation in animals and humans}.
\newblock Harvard University Press, 2000.

\bibitem{johnson2003puzzle}
D.~D.~P. Johnson, P.~Stopka, and S.~Knights.
\newblock The puzzle of human cooperation.
\newblock {\em Nature}, 421(6926):911--912, 2003.

\bibitem{rand2013human}
D.~G. Rand and M.~A. Nowak.
\newblock Human cooperation.
\newblock {\em Trends in cognitive sciences}, 17(8):413--425, 2013.

\bibitem{puurtinen2009between}
M.~Puurtinen and T.~Mappes.
\newblock Between-group competition and human cooperation.
\newblock {\em Proceedings of the Royal Society B: Biological Sciences}, 276(1655):355--360, 2009.

\bibitem{Borenstein13}
D.~B. Borenstein, Y.~Meir, J.~W. Shaevitz, and N.~S. Wingreen.
\newblock {{N}on-local interaction via diffusible resource prevents coexistence of cooperators and cheaters in a lattice model}.
\newblock {\em PLoS One}, 8(5):e63304, 2013.

\bibitem{Cremer19}
J.~Cremer, A.~Melbinger, K.~Wienand, T.~Henriquez, H.~Jung, and E.~Frey.
\newblock {{C}ooperation in {M}icrobial {P}opulations: {T}heory and {E}xperimental {M}odel {S}ystems}.
\newblock {\em J Mol Biol}, 431(23):4599--4644, Nov 2019.

\bibitem{Wingreen06}
N.~S. Wingreen and S.~A. Levin.
\newblock {{C}ooperation among microorganisms}.
\newblock {\em PLoS Biol}, 4(9):e299, Sep 2006.

\bibitem{Hamilton64a}
W.~D. Hamilton.
\newblock {{T}he genetical evolution of social behaviour. {I}}.
\newblock {\em J Theor Biol}, 7(1):1--16, Jul 1964.

\bibitem{Hamilton64b}
W.~D. Hamilton.
\newblock {{T}he genetical evolution of social behaviour. {I}{I}}.
\newblock {\em J Theor Biol}, 7(1):17--52, Jul 1964.

\bibitem{ohtsuki2006evolutionary}
H.~Ohtsuki and M.~A. Nowak.
\newblock Evolutionary games on cycles.
\newblock {\em Proceedings of the Royal Society B: Biological Sciences}, 273(1598):2249--2256, 2006.

\bibitem{ohtsuki2006simple}
H.~Ohtsuki, C.~Hauert, E.~Lieberman, and M.~A. Nowak.
\newblock A simple rule for the evolution of cooperation on graphs and social networks.
\newblock {\em Nature}, 441(7092):502--505, 2006.

\bibitem{chuang2009simpson}
J.~S. Chuang, O.~Rivoire, and S.~Leibler.
\newblock Simpson's paradox in a synthetic microbial system.
\newblock {\em science}, 323(5911):272--275, 2009.

\bibitem{Melbinger10}
A.~Melbinger, J.~Cremer, and E.~Frey.
\newblock {{E}volutionary game theory in growing populations}.
\newblock {\em Phys Rev Lett}, 105(17):178101, Oct 2010.

\bibitem{Cremer12}
J.~Cremer, A.~Melbinger, and E.~Frey.
\newblock {{G}rowth dynamics and the evolution of cooperation in microbial populations}.
\newblock {\em Sci Rep}, 2:281, 2012.

\bibitem{Kay20}
T.~Kay, L.~Keller, and L.~Lehmann.
\newblock {{T}he evolution of altruism and the serial rediscovery of the role of relatedness}.
\newblock {\em Proc Natl Acad Sci U S A}, 117(46):28894--28898, Nov 2020.

\bibitem{doebeli2005}
M.~Doebeli and C.~Hauert.
\newblock Models of cooperation based on the prisoner's dilemma and the snowdrift game.
\newblock {\em Ecology Letters}, 8(7):748--766, 2005.

\bibitem{Turner99}
P.~E. Turner.
\newblock Prisoner's dilemma in an {RNA} virus.
\newblock {\em Nature}, 398:441--443, 1999.

\bibitem{Gore09}
J.~Gore, H.~Youk, and A.~van Oudenaarden.
\newblock {{S}nowdrift game dynamics and facultative cheating in yeast}.
\newblock {\em Nature}, 459(7244):253--256, May 2009.

\bibitem{sigmund1999evolutionary}
K.~Sigmund and M.~A. Nowak.
\newblock Evolutionary game theory.
\newblock {\em Current Biology}, 9(14):R503--R505, 1999.

\bibitem{weibull1997evolutionary}
J.~W. Weibull.
\newblock {\em Evolutionary game theory}.
\newblock MIT press, 1997.

\bibitem{traulsen2009stochastic}
A.~Traulsen and C.~Hauert.
\newblock Stochastic evolutionary game dynamics.
\newblock {\em Reviews of nonlinear dynamics and complexity}, 2:25--61, 2009.

\bibitem{hummert2014evolutionary}
S.~Hummert, K.~Bohl, D.~Basanta, A.~Deutsch, S.~Werner, G.~Theissen, A.~Schroeter, and S.~Schuster.
\newblock Evolutionary game theory: cells as players.
\newblock {\em Molecular BioSystems}, 10(12):3044--3065, 2014.

\bibitem{lieberman2005evolutionary}
E.~Lieberman, C.~Hauert, and M.~A. Nowak.
\newblock Evolutionary dynamics on graphs.
\newblock {\em Nature}, 433(7023):312--316, 2005.

\bibitem{marrec2021toward}
L.~Marrec, I.~Lamberti, and A.-F. Bitbol.
\newblock Toward a universal model for spatially structured populations.
\newblock {\em Physical Review Letters}, 127(21):218102, 2021.

\bibitem{Abbara23}
A.~Abbara and A.-F. Bitbol.
\newblock Frequent asymmetric migrations suppress natural selection in spatially structured populations.
\newblock {\em PNAS Nexus}, 2(11):pgad392, 2023.

\bibitem{Houchmandzadeh11}
B.~Houchmandzadeh and M.~Vallade.
\newblock The fixation probability of a beneficial mutation in a geographically structured population.
\newblock {\em New Journal of Physics}, 13(7):073020, Jul 2011.

\bibitem{yagoobi2021fixation}
S.~Yagoobi and A.~Traulsen.
\newblock Fixation probabilities in network structured meta-populations.
\newblock {\em Scientific Reports}, 11(1):17979, 2021.

\bibitem{yagoobi2023}
S.~Yagoobi, N.~Sharma, and A.~Traulsen.
\newblock Categorizing update mechanisms for graph-structured metapopulations.
\newblock {\em Journal of The Royal Society Interface}, 20(200):20220769, 2023.

\bibitem{Kaveh15}
K.~Kaveh, N.~L. Komarova, and M.~Kohandel.
\newblock The duality of spatial death-birth and birth-death processes and limitations of the isothermal theorem.
\newblock {\em Royal Society Open Science}, 2(4):140465, 2015.

\bibitem{Pattni15}
K.~Pattni, M.~Broom, J.~Rychtář, and L.~J. Silvers.
\newblock Evolutionary graph theory revisited: when is an evolutionary process equivalent to the {M}oran process?
\newblock {\em Proceedings of the Royal Society A: Mathematical, Physical and Engineering Sciences}, 471(2182):20150334, 2015.

\bibitem{Hindersin15}
L.~Hindersin and A.~Traulsen.
\newblock Most undirected random graphs are amplifiers of selection for birth-death dynamics, but suppressors of selection for death-birth dynamics.
\newblock {\em PLOS Computational Biology}, 11(11):1--14, 11 2015.

\bibitem{tkadlec2020limits}
J.~Tkadlec, A.~Pavlogiannis, K.~Chatterjee, and M.~A. Nowak.
\newblock Limits on amplifiers of natural selection under death-birth updating.
\newblock {\em PLoS computational biology}, 16(1):e1007494, 2020.

\bibitem{Wallace75}
B.~Wallace.
\newblock {Hard and soft selection revisited}.
\newblock {\em Evolution}, 29(3):465--473, Sep 1975.

\bibitem{Slatkin81}
M.~Slatkin.
\newblock {{F}ixation probabilities and fixation times in a subdivided population}.
\newblock {\em Evolution}, 35(3):477--488, 1981.

\bibitem{moran1958random}
P.~A.~P. Moran.
\newblock Random processes in genetics.
\newblock In {\em Mathematical proceedings of the {C}ambridge philosophical society}, volume~54, pages 60--71. Cambridge University Press, 1958.

\bibitem{ewens2004mathematical}
W.~J. Ewens.
\newblock {\em Mathematical population genetics: theoretical introduction}, volume~1.
\newblock Springer, 2004.

\bibitem{Traulsen08}
A.~Traulsen, N.~Shoresh, and M.~A. Nowak.
\newblock {{A}nalytical results for individual and group selection of any intensity}.
\newblock {\em Bull Math Biol}, 70(5):1410--1424, Jul 2008.

\bibitem{gillespie1976general}
D.~T. Gillespie.
\newblock A general method for numerically simulating the stochastic time evolution of coupled chemical reactions.
\newblock {\em Journal of computational physics}, 22(4):403--434, 1976.

\bibitem{gillespie1977exact}
D.~T. Gillespie.
\newblock Exact stochastic simulation of coupled chemical reactions.
\newblock {\em The journal of physical chemistry}, 81(25):2340--2361, 1977.

\bibitem{Lenski91}
R.~E. Lenski, M.~R. Rose, S.~C. Simpson, and Tadler~S. C.
\newblock Long-term experimental evolution in \textit{Escherichia coli}. i. adaptation and divergence during 2,000 generations.
\newblock {\em Am. Nat.}, 138(6):1315--1341, 1991.

\bibitem{Elena03}
F.~E. Santiago and R.~E. Lenski.
\newblock Evolution experiments with microorganisms: the dynamics and genetic bases of adaptation.
\newblock {\em Nature Reviews Genetics}, 4(6):457--469, 2003.

\bibitem{Good17}
B.~H. Good, M.~J. McDonald, J.~E. Barrick, R.~E. Lenski, and M.~M. Desai.
\newblock The dynamics of molecular evolution over 60,000 generations.
\newblock {\em Nature}, 551(7678):45--50, 2017.

\bibitem{Kryazhimskiy12}
S.~Kryazhimskiy, D.~P. Rice, and M.~M. Desai.
\newblock {{P}opulation subdivision and adaptation in asexual populations of {S}accharomyces cerevisiae}.
\newblock {\em Evolution}, 66(6):1931--1941, Jun 2012.

\bibitem{Nahum15}
J.~R. Nahum, P.~Godfrey-Smith, B.~N. Harding, J.~H. Marcus, J.~Carlson-Stevermer, and B.~Kerr.
\newblock {{A} tortoise-hare pattern seen in adapting structured and unstructured populations suggests a rugged fitness landscape in bacteria}.
\newblock {\em Proc Natl Acad Sci U S A}, 112(24):7530--7535, Jun 2015.

\bibitem{France19}
M.~T. France and L.~J. Forney.
\newblock {{T}he {R}elationship between {S}patial {S}tructure and the {M}aintenance of {D}iversity in {M}icrobial {P}opulations}.
\newblock {\em Am Nat}, 193(4):503--513, 04 2019.

\bibitem{Chen20}
P.~Chen and R.~Kassen.
\newblock {{T}he evolution and fate of diversity under hard and soft selection}.
\newblock {\em Proc Biol Sci}, 287(1934):20201111, Sep 2020.

\bibitem{Kassen}
P.~P. Chakraborty, L.~R. Nemzer, and R.~Kassen.
\newblock {Experimental evidence that network topology can accelerate the spread of beneficial mutations}.
\newblock {\em Evolution Letters}, page qrad047, 2023.

\bibitem{Erez20}
A.~Erez, J.~G. Lopez, B.~G. Weiner, Y.~Meir, and N.~S. Wingreen.
\newblock {{N}utrient levels and trade-offs control diversity in a serial dilution ecosystem}.
\newblock {\em Elife}, 9, Sep 2020.

\bibitem{Chen08}
Xiaojie Chen, Feng Fu, and Long Wang.
\newblock Influence of different initial distributions on robust cooperation in scale-free networks: A comparative study.
\newblock {\em Physics Letters A}, 372(8):1161--1167, 2008.

\bibitem{Zhang15}
Y.~Zhang, F.~Fu, X.~Chen, G.~Xie, and L.~Wang.
\newblock {{C}ooperation in group-structured populations with two layers of interactions}.
\newblock {\em Sci Rep}, 5:17446, Dec 2015.

\bibitem{blyth1972simpson}
C.~R. Blyth.
\newblock On {S}impson's paradox and the sure-thing principle.
\newblock {\em Journal of the American Statistical Association}, 67(338):364--366, 1972.

\bibitem{simpson1951interpretation}
E.~H. Simpson.
\newblock The interpretation of interaction in contingency tables.
\newblock {\em Journal of the Royal Statistical Society: Series B (Methodological)}, 13(2):238--241, 1951.

\bibitem{samuels1993simpson}
M.~L. Samuels.
\newblock Simpson's paradox and related phenomena.
\newblock {\em Journal of the American Statistical Association}, 88(421):81--88, 1993.

\bibitem{good1987amalgamation}
I.~J. Good and Y.~Mittal.
\newblock The amalgamation and geometry of two-by-two contingency tables.
\newblock {\em The Annals of Statistics}, pages 694--711, 1987.

\bibitem{Fu13}
F.~Fu and M.~A. Nowak.
\newblock {{G}lobal migration can lead to stronger spatial selection than local migration}.
\newblock {\em J Stat Phys}, 151(3-4):637--653, May 2013.

\bibitem{Wu11}
T.~Wu, F.~Fu, and L.~Wang.
\newblock {{M}oving away from nasty encounters enhances cooperation in ecological prisoner's dilemma game}.
\newblock {\em PLoS One}, 6(11):e27669, 2011.

\bibitem{CrowKimura}
J.~F. Crow and M.~Kimura.
\newblock {\em {An Introduction to Population Genetics Theory}}.
\newblock Blackburn, 2009.

\bibitem{kimura64}
M.~Kimura.
\newblock Diffusion models in population genetics.
\newblock {\em Journal of Applied Probability}, 1(2):177–232, 1964.

\bibitem{Blythe2007}
R.~A. Blythe and A.~J. McKane.
\newblock Stochastic models of evolution in genetics, ecology and linguistics.
\newblock {\em Journal of Statistical Mechanics: Theory and Experiment}, 2007(07):P07018, jul 2007.

\bibitem{Ewens79}
W.~J. Ewens.
\newblock {\em {Mathematical Population Genetics}}.
\newblock Springer-Verlag, 1979.

\end{thebibliography}

\end{document}